\documentclass[useAMS,usenatbib]{mn2e}
\usepackage{epsfig}
\usepackage{epstopdf}
\usepackage{graphicx}
\usepackage{amsmath}
\usepackage{appendix}
\usepackage[T1]{fontenc}
\usepackage{ae,aecompl}

\usepackage{graphicx}	% Including figure files
\usepackage{amsmath}	% Advanced maths commands
\usepackage{amssymb}	% Extra maths symbols

\title[Crossing Curved Light Surfaces]{An Improved Algorithm for Crossing Curved Light Surfaces:
rapidly rotating pulsar magnetospheres in curved spacetime}

\author[Huang et al.]{
Lei Huang$^{1}$\thanks{muduri@shao.ac.cn},
Zhen Pan$^{2}$\thanks{zhpan@ucdavis.edu},
Cong Yu$^{3}$\thanks{yucong@mail.sysu.edu.cn}
\\
$^{1}$Key Laboratory for Research in Galaxies and Cosmology, \\Shanghai Astronomical Observatory, Chinese Academy of Sciences,
Shanghai, 200030, China\\
$^{2}$Department of Physics, University of California, Davis, CA, 95616, USA\\
$^{3}$School of Physics and Astronomy, Sun Yat-sen University, Zhuhai 519082, China
}

\date{Accepted XXX. Received YYY; in original form ZZZ}

\pubyear{2018}

\begin{document}
\label{firstpage}
\pagerange{\pageref{firstpage}--\pageref{lastpage}}
\maketitle

\begin{abstract}
The structure of  force-free, steady and axisymmetric magnetosphere of a neutron star (NS) is governed by the Grad-Shafranov (GS) equation, which is a second-order differential equation but degrades to first-order on the light surface (LS). The key to numerically solving the GS equation is to enable magnetic field lines smoothly cross the LS, and crossing a straight LS in flat spacetime has been a well-studied problem. But the numerical algorithm implementation becomes complicate in the presence of a bent LS, e.g. in curved spacetime, since there is no suitable computation grid adapted to it. We propose to circumvent this grid-LS mismatch problem by introducing a new coordinate frame designed such that the LS in it is a straight line. As an application, we investigate the general relativistic (GR) effect in magnetosphere structure of rapidly rotating pulsars in detail, where the LS is bent towards the central NS. We split the GR effect into two parts, curvature and frame-dragging; measure each of them and examine their dependence on the NS mass and the angular velocity for pulsars embedded in aligned dipole and multipole magnetic fields. Qualitatively speaking, we find that the curvature effect compactifies  the magnetic field lines near the NS, therefore reduces the open magnetic flux and the Poynting luminosity, while the frame-dragging effect contributes a minor part in shaping the magnetosphere structure but plays a role in enhancing the spacelike current generation.
\end{abstract}

\begin{keywords}
gravitation -- relativistic processes -- magnetic fields
\end{keywords}

\section{Introduction}
The magnetosphere structure of compact objects has been one of most important topics in astrophysics since decades ago. In their seminal work, \citet{GJ69} outlined the basic physics of  neutron star (NS) magnetospheres.
Since then a large amount of works have been done to investigate the magnetosphere global structure
\citep[e.g.][]{Sturrock71,Ruderman1972,RS75,Arons1979,Michel1982, Arons1983, Contop99,Gruz05,Timo06}. Now the community has reached to a consensus that a magnetosphere is highly magnetically dominated,
therefore is force-free to a good approximation {\em if} there exists a plasma that is dense enough to screen the electric field parallel to the magnetic field.

With the force-free approximation, the magnetosphere structure is
governed by the Grad-Shafranov (GS) equation, which is a second-order differential equation of $\phi$ component of the electromagnetic vector potential
$\Psi(\vec x)$ with to be determined poloidal electric current $I(\Psi)$ flowing along the magnetic field lines. The GS equation degrades to first-order on the light surface (LS), where the rotation speed of the field lines approaches light speed, therefore plasma particles stop corotating with the field lines. Consequently, there is no closed field lines extend beyond the LS, and there is inevitably a non-zero particle flow along each open field line extending from the star surface to infinity. As  proposed by \citet{Contop99}, the key to numerically solving the GS equation is to enable the field lines smoothly cross the LS by adjusting the current  $I(\Psi)$. The numerical algorithm has been well studied for pulsar magnetospheres in flat spacetime, and has been improved to  percent level precision in terms of the Poynting luminosity \citep[e.g.][]{Gruz05,Timo06}.

There has been longstanding uncertainties about the source of plasma and the pulsar emission mechanism  in the above picture of force-free magnetospheres. The charged particles would be depleted along the open field lines if there was no efficient refilling mechanism. A well accepted refilling mechanism is pair production: emission of gamma-rays from accelerated particles in the magnetic field and subsequent conversion to $e^{\pm}$ pairs via photon-magnetic field collision and photon-photon collision  \citep[e.g.][]{Sturrock71,Cheng1986,Bai2010, Petri2016b, Cerutti2017,Venter2017,Philippov2018}. As shown in several particle-in-cell simulations \citep[e.g.][]{Chen2014,Philippov2014,Philippov2015a}, the magnetosphere relaxes to the force-free state, {\em if} there is sufficient pair production extending  from the NS  surface to the LS, while the magnetosphere settles into the electrosphere state \citep{Jackson1976,Krause1985} if the pair  production is confined within a much limited region. Another uncertainty in pulsar physics is the emission mechanism, e.g. curvature radiation \citep[e.g.][]{Sturrock71,RS75}, and inverse Compton scattering by secondary relativistic particles \citep[e.g.][]{Qiao88,Lv11}, are also closely  connected to the pair production.
Therefore it is crucial to figure out the regions of pair production for testing the self-consistence of
force-free magnetospheres. Whereas kinetic simulation from first principle is difficult,  a simple prescription was summarized from previous kinetic simulations that  pair production takes place where the electric current is spacelike  \citep{Beloborodov2008,Timokhin2013,Philippov2015,Belyaev2016}.
Taking advantage of this  prescription, one can find out the pair production regions from the force-free magnetospheres without invoking underlying physics of gamma-ray emission and the subsequent conversion to $e^\pm$ pairs. Based on this prescription, recent analytic studies \citep{Gralla16, Belyaev2016} showed that  general relativistic (GR) effect plays an important role in generating spacelike current and therefore in pair production.

Other than its significance in the pair production,  it is of interest to investigate how much the GR effect changing the structure of pulsar magnetospheres \citep[e.g.][]{Palenzuela2013,Ruiz14,Philippov2015,Petri16}, including the configuration of magnetic field lines, the current flow and the Poynting luminosity,  due to the strong gravity around NSs ($GM/c^2/r_{\rm NS}\sim 10\%$, with $M$ and $r_{\rm NS}$ being the mass and the radius of a typical NS).

In this paper, we present the first work of investigating the GR effect in the pulsar magnetospheres by numerically solving the GS equation in curved spacetime.
 We systematically examine the GR corrections dependence on the properties of the central NS, specifically on the angular velocity and the mass (compactness). In flat spacetime, the  LS is away from the rotation axis by a constant distance, while the LS is bent towards the central star in curved spacetime. In principle, we can use the same numerical algorithm as in flat spacetime. But the the LS is bent and the usual computation grid points scatter around the LS, consequently we can only determine the $\Psi$ values on LS by extrapolation from grid points nearby, which complicates the algorithm and likely sacrifices some numerical accuracy. To avoid these problems, we propose to introduce a new coordinate frame designed such that the curved-spacetime-LS in it is a straight line. Then it is straightforward implement the usual algorithm, and we expect a similar percent-level numerical precision in term of the Poynting luminosity, though the form of the GS equation in the new coordinate system becomes a bit cumbersome due to extra terms arising from coordinate transformation.

This paper is organized as follows. We explain our numerical algorithm in Section \ref{sec:algorithm}, and systematically explore how the magnetosphere structure of dipole field and multipole field impacted by the GR effect in Section \ref{sec:GR} and Section \ref{sec:GRm}, respectively. Summary and discussion are given in Section \ref{sec:summary}. For reference, we also list some details of the numerical algorithm in Appendices. Throughout this paper, we adopt the natural units $G=c=1$.

\section{Numerical Algorithm}
\label{sec:algorithm}

We consider a millisecond pulsar with angular velocity $\Omega$ and mass $M$.
In Boyer-Linquist (BL) coordinates, the spacetime outside the NS is described by
the ``Kerr" metric \citep{HT68}
\begin{equation}
    ds^2= \alpha^2 dt^2 + \alpha^{-2} dr^2 + r^2 d\theta^2   + r^2\sin^2\theta (d\phi - \Omega_Z dt)^2\ ,
\end{equation}
where $\alpha^2=1-2M/r$  and $\Omega_Z=2\Omega\hat{I}_{\rm NS}/r^3$
is the frequency of frame-dragging, with $\hat{I}_{\rm NS}=(2/5)\cdot Mr_{\rm NS}^2$ being
the moment of inertia  for a rigid rotating and uniform-density star.

The force-free, axisymmetric and steady magnetosphere outside the NS is governed by the curved-spacetime GS equation
\citep[e.g.][]{Gralla14,Gralla16}, which we write in the symmetric form proposed by \citet{Pan17},
\begin{eqnarray}
\label{eq:GS_1}
    0&=& \left( \Psi_{,rr} + \frac{\sin^2\theta}{\alpha^2r^2}  \Psi_{,\mu\mu}  \right)\ \mathcal{K}(r,\theta;\Omega) \nonumber\\
    &+& \left(  \Psi_{,r}\ \partial_r + \frac{\sin^2\theta}{\alpha^2r^2} \Psi_{,\mu} \partial_\mu \right)\ \mathcal{K}(r,\theta;\Omega) \ +\ \frac{II'(\Psi)}{\alpha^2} \ ,
\end{eqnarray}
where $\Psi(r,\theta)$ is the toroidal component of the vector potential, $\mu=\cos\theta$, $I(\Psi)$ is the poloidal current flowing along the magnetic field lines, and we have defined the LS function
\begin{eqnarray}
\label{eq:LS_func}
    \mathcal{K}(r,\theta;\Omega) =  \alpha^2 - r^2 \sin^2\theta\ (\Omega-\Omega_Z)^2 ,
\end{eqnarray}
which is prefactor of the second-order differential terms.
The GS equation degrades to first-order on the LS, where  $\mathcal{K}(r,\theta;\Omega)=0$.

In the remaining part of this paper, we explore the GR effect for millisecond pulsars
in the parameter space $M\in [0, 2M_\odot]$, $r_{\rm NS} = 10~{\rm km}$, and $\Omega r_{\rm NS} \in \{0.1, 0.2\}$,
among which we select three fiducial millisecond pulsars as our benchmarks (see Table \ref{table:1}).

\begin{table}
\caption{Three fiducial millisecond pulsars we explored in detail.}
\label{table:1}
\begin{tabular}{ |c|c|c|c}
      & $M(M_\odot)$ & $r_{\rm NS}$(km) & $\Omega r_{\rm NS} $ \\
 \hline
 Pulsar 1 & $1$  & $10$  & $0.1$ \\
 Pulsar 2 & $1$  & $10$  & $0.2$ \\
 Pulsar 3 & $2$  & $10$  & $0.2$ \\
 \hline
\end{tabular}
\end{table}

\subsection{New Coordinates}

In  flat spacetime $\alpha=1$ and $\Omega_Z=0$, GS equation (\ref{eq:GS_1}) reduces to the well-known form of pulsar equation \citep[e.g.][]{Scharlemann1973} and the corresponding LS is located at $r\sin\theta= 1/\Omega \equiv R_{\rm LS}$, which is a straight line in the cylindrical coordinates $(R= r\sin\theta, Z=r\cos\theta)$. As proposed by \citet{Contop99}, one can construct the global structure of the magnetosphere by numerically solving the GS equation in the interior region ($R < R_{\rm LS}$) and in the outer region ($R > R_{\rm LS}$) separately, and match the field lines on the LS by requiring $\Psi(R_{\rm LS}^+) = \Psi(R_{\rm LS}^-)$ via adjusting poloidal current $I(\Psi)$.

\begin{figure*}
    \includegraphics[scale=0.38]{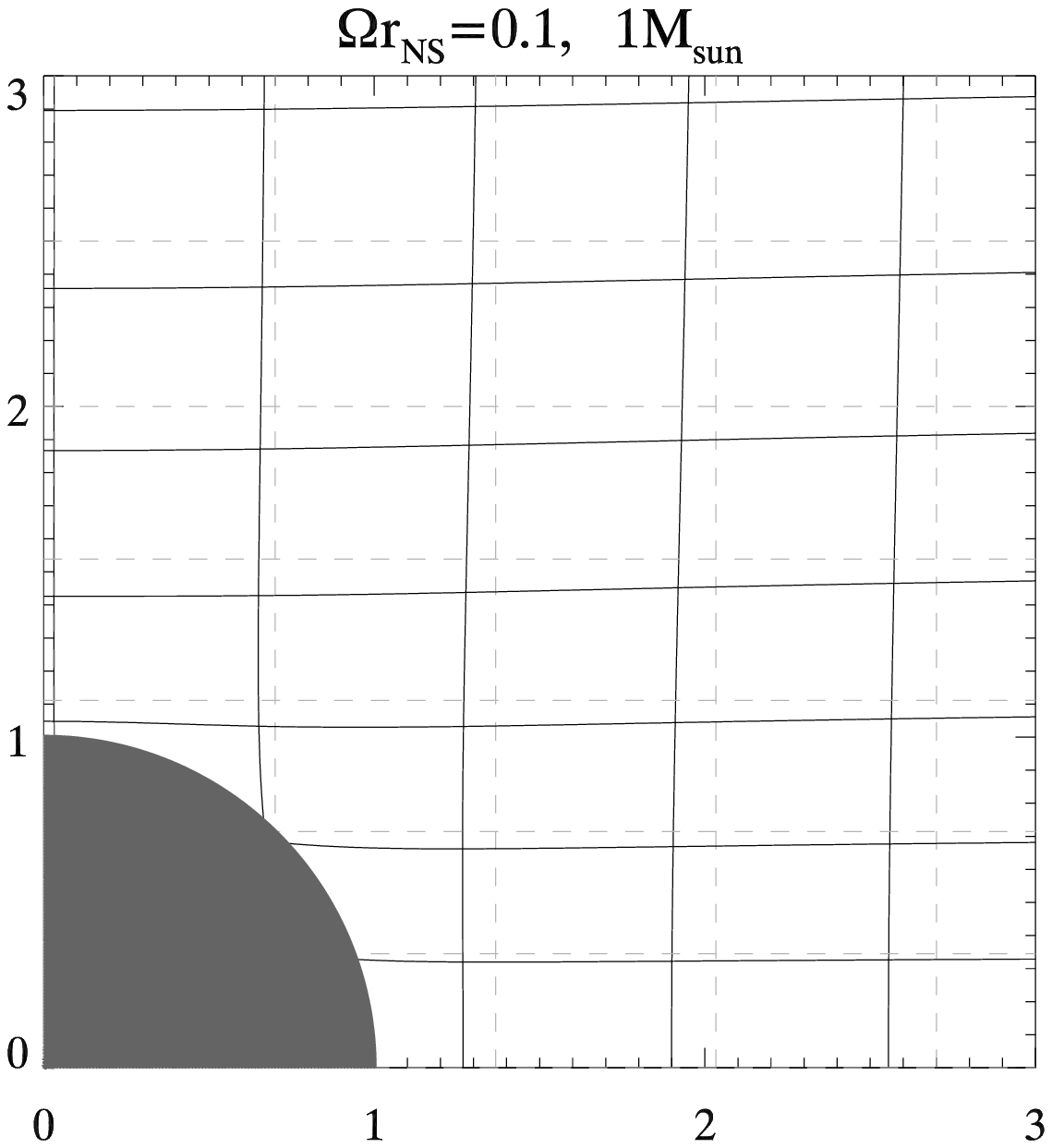}
    \includegraphics[scale=0.38]{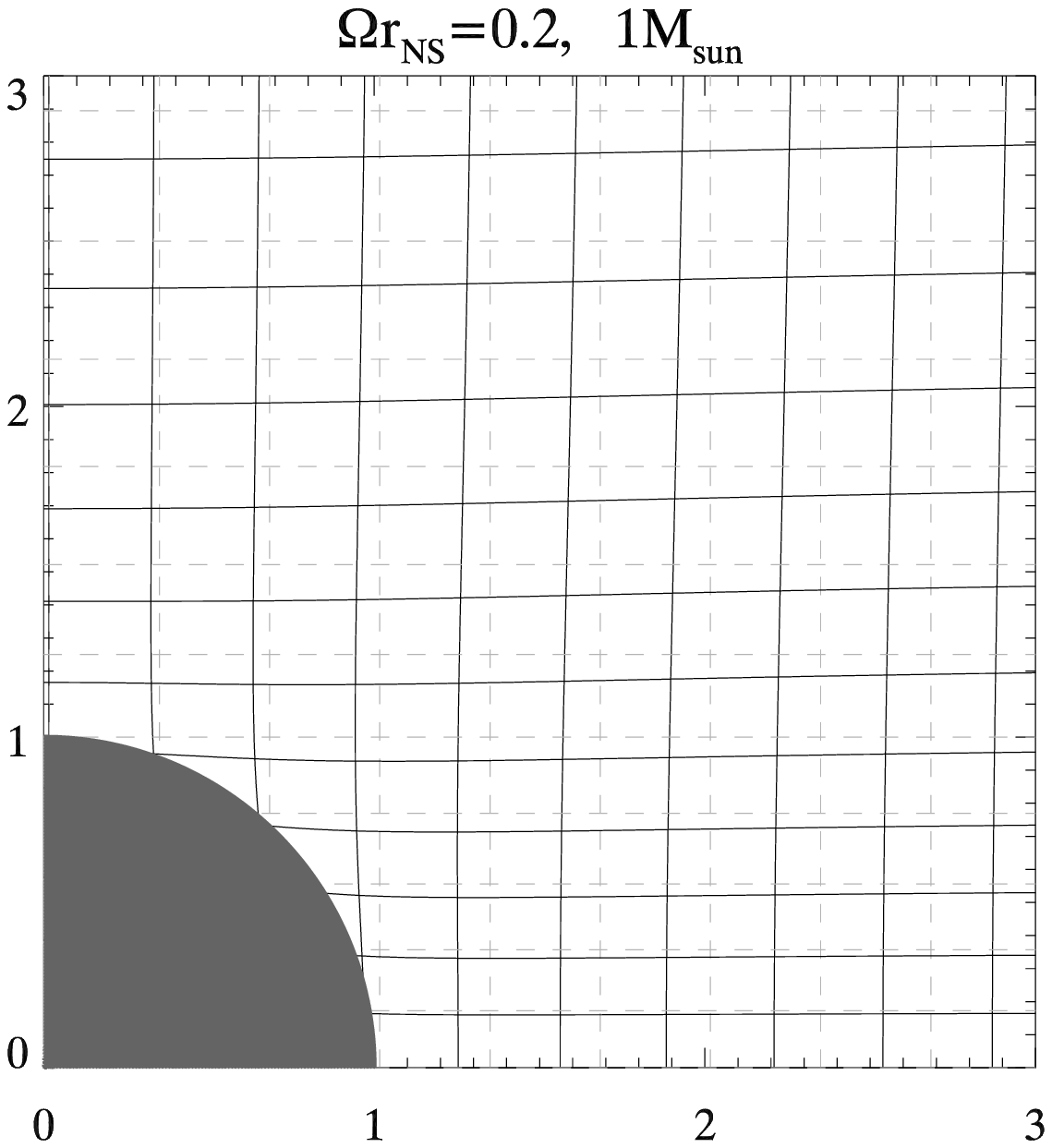}
    \includegraphics[scale=0.38]{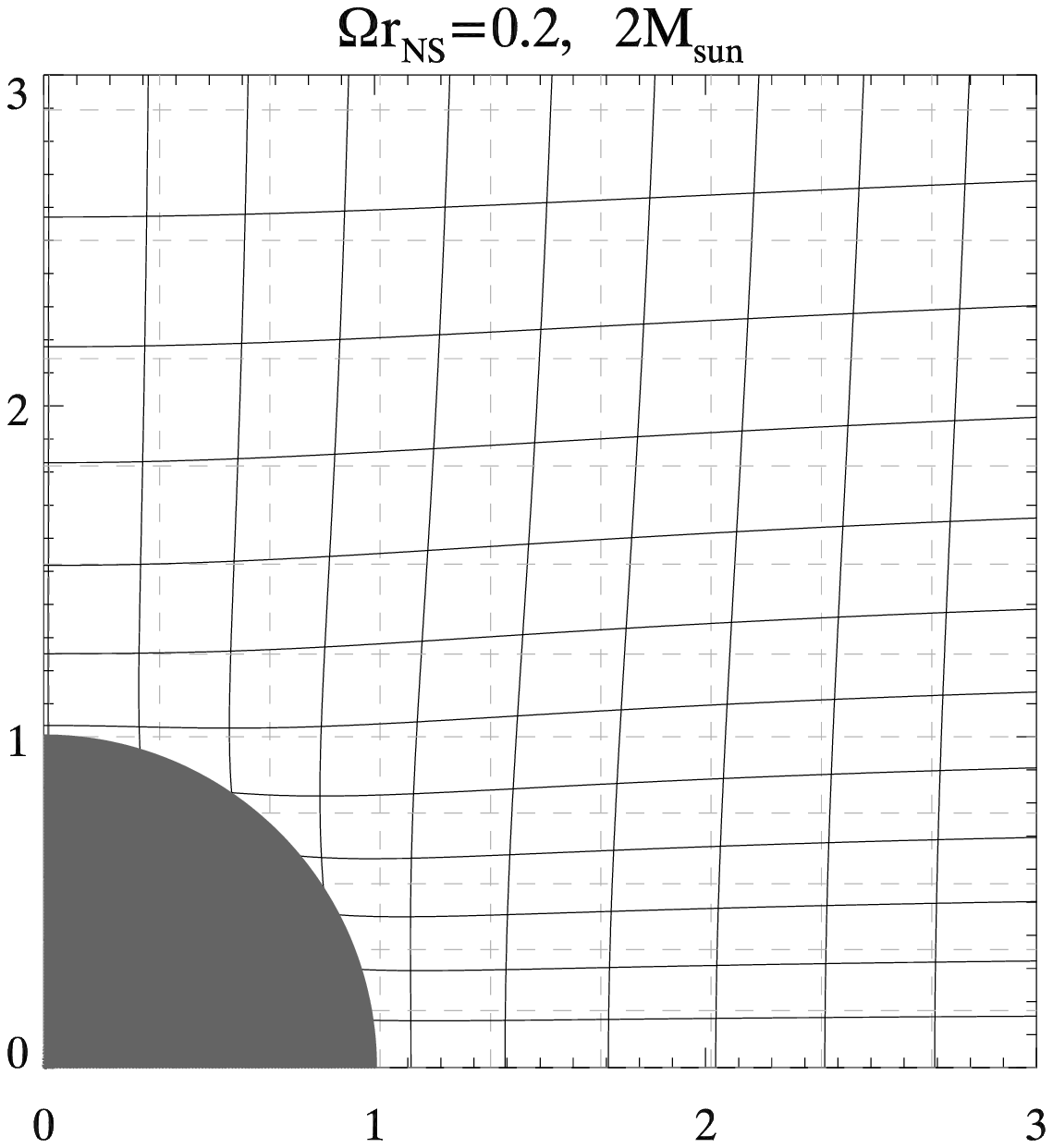}
\caption{\label{fig:Grid}
Uniform $(\tilde{R},\tilde{Z})$ grid plotted in the  $\tilde R-\tilde Z$ plane ({\em dashed grey}) and  projected onto the $R-Z$ plane ({\em solid black}) for the three fiducial pulsars.
 }
\end{figure*}

\begin{figure*}
    \includegraphics[scale=0.38]{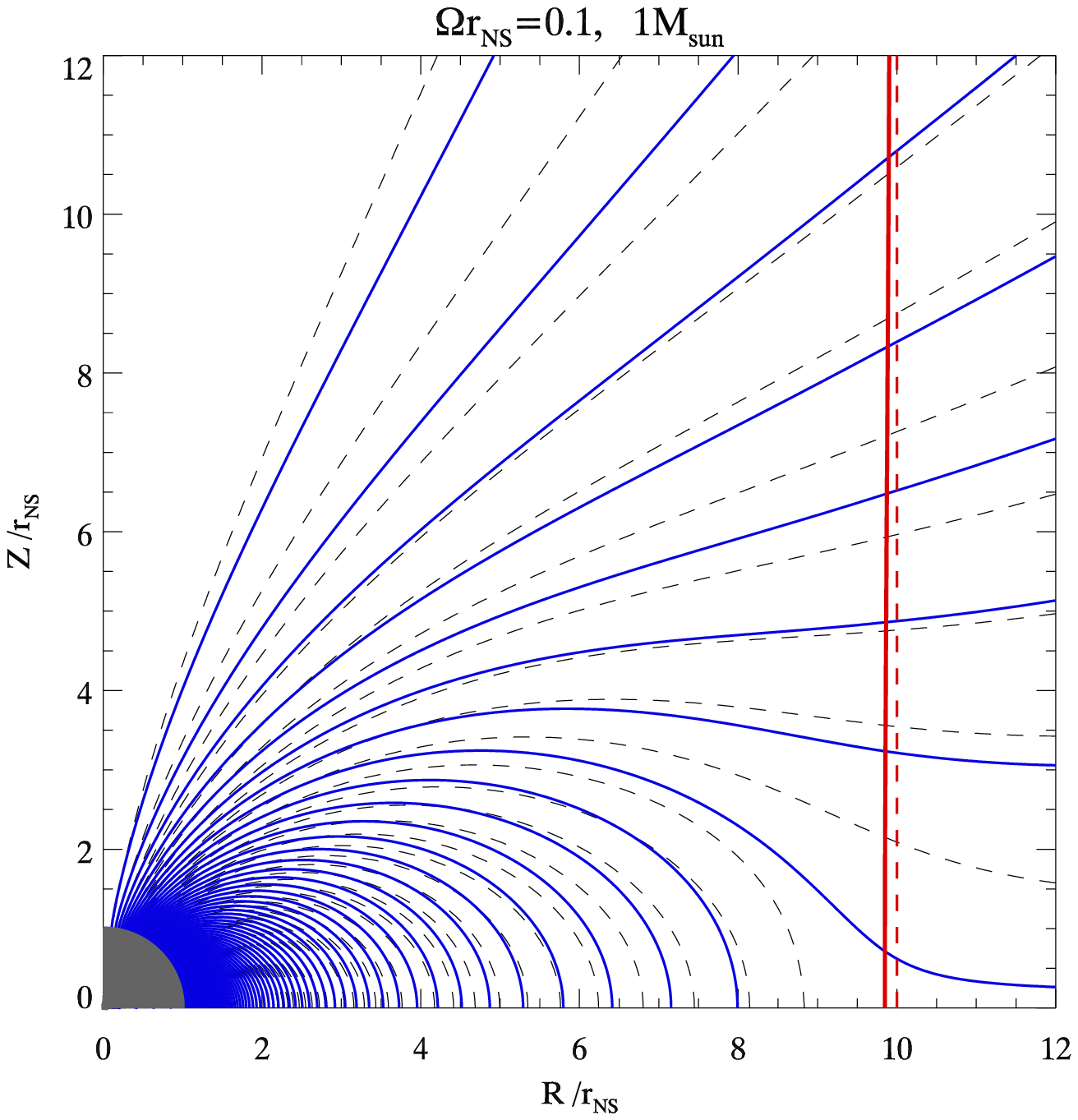}
    \includegraphics[scale=0.38]{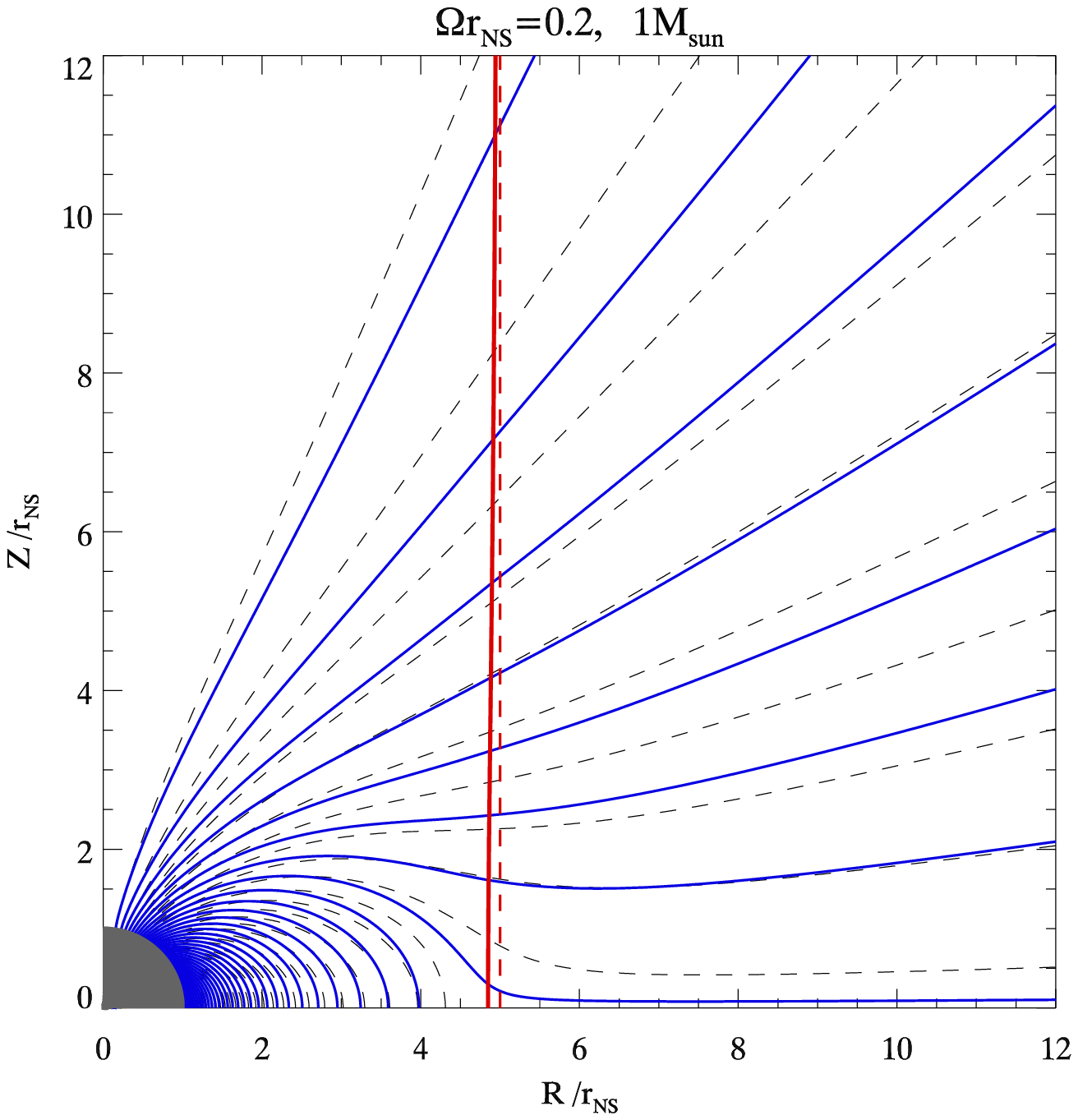}
    \includegraphics[scale=0.38]{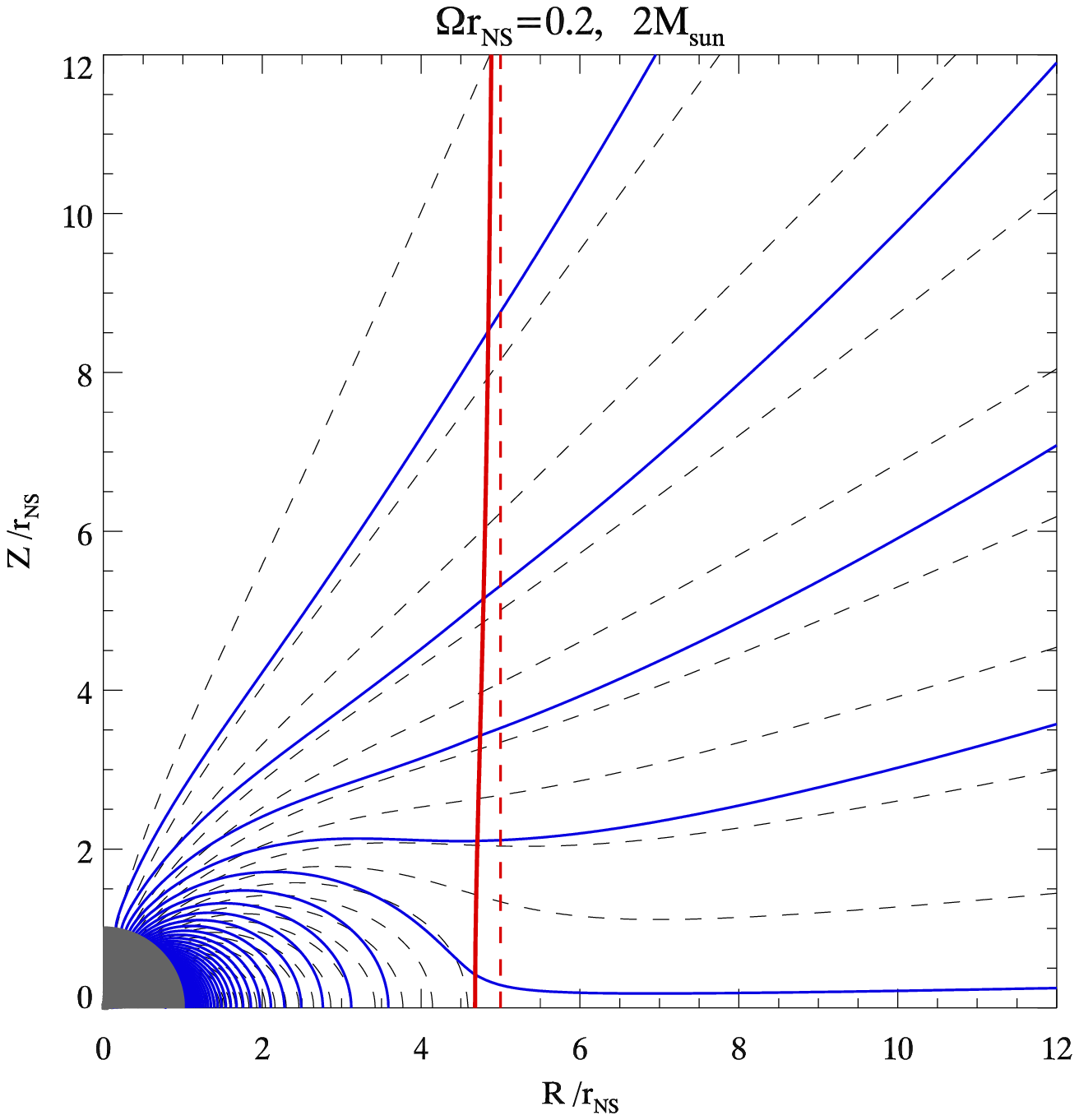}
\caption{\label{fig:NS02}
The field-line configurations of fiducial millisecond Pulsar 1 ({\em Left}), Pulsar 2 ({\em Middle}),
and Pulsar 3 ({\em Right}) in Kerr spacetime ({\rm solid blue}) and in flat spacetime ({\rm dashed grey}).
In each panel, the solid and dashed red lines represent LS in Kerr and flat spacetime, respectively.
The LS in Kerr spacetime is bent towards the central NS by $\sim M/R_{\rm LS} \approx 1.5\%, 3\%, 6\%$
on the equatorial plane for Pulsar 1, 2, \& 3, respectively.}
\end{figure*}

However the LS is bent towards the central NS in  curved spacetime (see Figure \ref{fig:NS02} or Equation \ref{eq:LS_func}), which makes the usual algorithm of matching field lines on the LS rather complicate, since there is no easy way to place computation grid points exactly on the bent LS. We find it is helpful to introduce a new coordinate frame in which the LS is a straight line, and the grid-LS mismatch complication
would be avoided. For this purpose, we perform a coordinate transformation by two steps. In the first step, we introduce a new set of spherical coordinates $(\tilde{r},\tilde{\theta})$, which relate to the old ones by
\begin{eqnarray}
    &\ & \tilde{r}= \frac{r}{\alpha\beta}\ , \ \tilde{\theta}= \theta\ ,
\end{eqnarray}
where
\begin{eqnarray}
    \beta&=& \left( 1 - \frac{\Omega_Z(r)}{\Omega} \right)^{-1} = \left( 1 - \frac{2\hat{I}_{\rm NS}}{r^3} \right)^{-1}\ ,
\end{eqnarray}
capturing the frame-dragging effect. In the ($\tilde r,\tilde \theta$) coordinates, we restore the LS function to
its flat spacetime form $\mathcal K = 1-\Omega^2\tilde r^2\sin^2\tilde \theta^2$.

In the next step, we introduce cylindrical coordinates $(\tilde R, \tilde Z)$ relating the spherical coordinates
 $(\tilde r, \tilde \theta)$ by
\begin{eqnarray}
    \tilde{R}= \tilde{r}\sin\tilde{\theta} \ ,\ \tilde{Z}= \tilde{r}\cos\tilde{\theta}\ .
\end{eqnarray}
In the new cylindrical coordinates, the LS function is written as $\mathcal K = 1-\Omega^2\tilde R^2$ and  the LS as expected is a straight line, given by $\tilde{R}=1/\Omega\equiv\tilde{R}_{\rm LS}$.
The GS equation turns to
\begin{eqnarray}
\label{EQ:GS_2}
    0&=& (1-\Omega^2\tilde{R}^2)\ \mathcal{D}\ \partial_{\tilde{R}\tilde{R}}\Psi + (1-\Omega^2\tilde{R}^2)\ \mathcal{E}\ \partial_{\tilde{Z}\tilde{Z}}\Psi  \nonumber\\
    &\ & + \left( \mathcal{F} - \frac{1+\Omega^2\tilde{R}^2}{\beta^2} \right) \  \frac{\partial_{\tilde{R}}\Psi}{\tilde{R}} + \mathcal{G} \ \frac{\partial_{\tilde{Z}}\Psi}{\tilde{Z}}  \nonumber\\
    &\ & + (1-\Omega^2\tilde{R}^2)\ \mathcal{H} \ \partial_{\tilde{R}\tilde{Z}} \Psi + II'(\Psi)\ ,
\end{eqnarray}
where the coefficients $(\mathcal{D,E,F,G,H})$ are given in detail in Appendix \ref{sec:appB}.
Note that the GR corrections $\alpha, \beta,\mathcal{D},\mathcal{E}\to1$ and
$\mathcal{F},\mathcal{G},\mathcal{H}\to0$ as $r\rightarrow \infty$.

To obtain some intuition of how much difference between the new coordinates and the old ones,
we plot a set of uniform $(\tilde{R},\tilde{Z})$ grid in the $\tilde R-\tilde Z$ plane
and also project it onto the $R-Z$ plane for the three fiducial pulsars in Figure \ref{fig:Grid}.

\subsection{Numerical Techniques}

For numerical convenience, we further introduce normalized coordinates
$(\tilde{x}=\Omega\tilde{R}, ~\tilde{z}=\Omega\tilde{Z})$, and normalized current  $\tilde{A}=I/\Omega$.
The GS equation is rewritten as
\begin{eqnarray}
\label{eq:GS_0}
    0&=& (1-\tilde{x}^2)\ \left( \mathcal{D}\ \partial_{\tilde{x}\tilde{x}}\Psi + \mathcal{E}\ \partial_{\tilde{z}\tilde{z}}\Psi + \frac{1}{\beta^2} \frac{\partial_{\tilde{x}}\Psi}{\tilde{x}} + \mathcal{H} \ \partial_{\tilde{x}\tilde{z}} \Psi \right)  \nonumber\\
    &\ & + \left( \mathcal{F} - \frac{2}{\beta^2} \right) \  \frac{\partial_{\tilde{x}}\Psi}{\tilde{x}} + \mathcal{G} \ \frac{\partial_{\tilde{z}}\Psi}{\tilde{z}}  + \tilde{A}\tilde{A}_{,\Psi}\ .
\end{eqnarray}
We use the method of Successive Overrelaxation (SOR) in Numerical Recipes \citep{NR} to solve the GS equation.

On the LS $\tilde{x}=1$, the second-order GS equation (\ref{eq:GS_0}) degrades to first-order, i.e.,
\begin{eqnarray}
    \partial_{\tilde{x}}\Psi&=& \left. \frac{1}{2/\beta^2-\mathcal{F}} \ \left( \mathcal{G} \ \frac{\partial_{\tilde{z}}\Psi}{\tilde{z}}  + \tilde{A}\tilde{A}_{,\Psi} \right) \right|_{\tilde{x}=1} \ ,
\end{eqnarray}
which requires special treatment other than the usual relaxation algorithm designed for evolving second-order partial differential equations.
For this purpose, we rewrite the GS equation as
\begin{eqnarray}
\label{eq:GS_LS}
    0&=& \partial_{\tilde{x}\tilde{x}} \Psi + \frac{\mathcal{E}}{\mathcal{D}}\  \partial_{\tilde{z}\tilde{z}} \Psi + \frac{\mathcal{H}}{\mathcal{D}}\  \partial_{\tilde{x}\tilde{z}} \Psi \\
    &\ & + \frac{1}{\mathcal{D}}\left\{  \frac{ f(\tilde{x}, \tilde{z}; \beta, \mathcal{F}) \partial_{\tilde{x}} \Psi + g(\tilde{x}, \tilde{z}; \mathcal{G}) \partial_{\tilde{z}} \Psi + \tilde{A}\tilde{A}_{,\Psi} }{(1-\tilde{x}^2)}   \right\} \ , \nonumber
\end{eqnarray}
and  we further remove the $0/0$ singularity using the L'H\^opital's rule
\begin{eqnarray}
    &\ & \lim_{\tilde{x}\to1} \frac{ f \partial_{\tilde{x}} \Psi + g \partial_{\tilde{z}} \Psi + \tilde{A}\tilde{A}_{,\Psi} }{(1-\tilde{x}^2)} \nonumber\\
    &=& -\frac{1}{2} \Bigg[ f \partial_{\tilde{x}\tilde{x}} \Psi + f_{\tilde{x}} ( \tilde{z}; \beta,\mathcal{F};\ \partial_{\tilde{x}} \beta, \partial_{\tilde{x}} \mathcal{F} )\  \partial_{\tilde{x}} \Psi    \nonumber\\
    &\ &  + g \partial_{\tilde{x}\tilde{z}} \Psi + g_{\tilde{x}} ( \tilde{z}; \mathcal{G}; \partial_{\tilde{x}} \mathcal{G} )\ \partial_{\tilde{z}} \Psi + \frac{{\rm d} (\tilde{A}\tilde{A}_{,\Psi})}{{\rm d}\Psi} \partial_{\tilde{x}}\Psi  \Bigg]\ ,
\end{eqnarray}
where functions $f$ and $g$ are
\begin{eqnarray}
    f(\tilde{x}, \tilde{z}; \beta, \mathcal{F}) = \frac{\mathcal{F}-(1+\tilde{x}^2)/\beta^2}{ \tilde{x}},
    \quad g(\tilde{x}, \tilde{z}; \mathcal{G}) = \frac{\mathcal{G}}{\tilde{z}}\ ,
\end{eqnarray}
and details of $\partial_{\tilde{x}} \beta, \partial_{\tilde{x}} \mathcal{F}, \partial_{\tilde{x}} \mathcal{G}$ are given in Appendix \ref{sec:appC}.

Practically, we use the SOR to  solve the GS equation (\ref{eq:GS_0}) in the outside region $\tilde x > 1$
and update the $\Psi$ values on the LS $\Psi(\tilde x_{\rm LS}^+)$  according to Equation (\ref{eq:GS_LS});
in the same way, we also solve the GS equation (\ref{eq:GS_0}) in the inside region $\tilde x < 1$
and update the $\Psi$ values on the LS $\Psi(\tilde x_{\rm LS}^-)$ according to Equation (\ref{eq:GS_LS}).
For a trial current $\tilde A(\Psi)$, we expect no agreement between $\Psi(\tilde x_{\rm LS}^+)$ and $\Psi(\tilde x_{\rm LS}^-)$. To ensure smooth field lines across the LS, we iteratively correct $\tilde A(\Psi)$ as follows \citep[e.g.][]{Contop99,Gruz05,Huang16},
\begin{eqnarray}
    (\tilde{A}\tilde{A}_{,\Psi})(\Psi_{\rm new}) &=& (\tilde{A}\tilde{A}_{,\Psi})(\Psi_{\rm old}) + \mu_1 [ \Psi(\tilde{x}_{\rm LS}^+) - \Psi(\tilde{x}_{\rm LS}^-) ]\ ,  \nonumber\\
    \Psi_{\rm new} &=& 0.5 [ \Psi(\tilde{x}_{\rm LS}^+) + \Psi(\tilde{x}_{\rm LS}^-) ]\ ,
\end{eqnarray}
where $\mu_1$ is chosen empirically. In each step of iteration,
the poloidal current $\tilde{A}(\Psi)$ is obtained by integrating $(\tilde{A}\tilde{A}_{,\Psi})$
from the pole to the last open field $\Psi_{\rm last}$. Usually $\tilde A(\Psi_{\rm last})$ does not vanish,
and there should be a return current sheet along the last open field line closing the current circuit.
Numerically we approximate the current sheet as a return current $\tilde{A}_{\rm ret}$ in a narrow range $[\Psi_{\rm last},\Psi_{\rm last}+\delta]$ in the form of $\tilde{A}_{\rm ret}(\Psi)=\tilde{A}_0\left[ ( \Psi - (\Psi_{\rm last} + \delta/2) )^2 - (\delta/2)^2 \right]$, where the constant $\tilde{A}_0$ is determined by
$\int_0^{\Psi_{\rm last}+\delta}\tilde{A}(\Psi)d\Psi=0$ to close the current circuit \citep{Timo06}, and $\delta$
is the numerical width of the return current sheet.

\section{GR Effect in Pulsar Magnetosphere Structure: Aligned Dipole}
\label{sec:GR}

For each NS specified by three parameters $\{M, r_{\rm NS}, \Omega\}$,
we numerically solve the GS equation in flat/Schwarzschild/Kerr spacetime, respectively.
In order to quantify the GR corrections,
we use the same boundary condition
\begin{eqnarray}
\label{eq:bc}
    \Psi(r_{\rm NS}, \theta) = B_1 r_{\rm NS}^2 \sin^2\theta,
\end{eqnarray}
for each case, with $ B_1$ being a constant.

In Figure \ref{fig:NS02}, we  plot  the field-line configurations of the three fiducial millisecond pulsars (Table \ref{table:1}),  which clearly shows that the field-line configuration is more compact in  Kerr spacetime (solid blue) than in flat spacetime (dashed grey), and therefore the open magnetic flux $\Psi_{\rm last}$ decreases in curved spacetime. In each panel, the solid and dashed red line represent the LS in Kerr spacetime and in flat spacetime, respectively. We find the LS  in Kerr spacetime is bent towards the central NS by $\sim  M/R_{\rm LS} \approx 1.5\%, 3\%,  6\%$ on the equatorial plane for Pulsar 1, 2, \& 3, respectively. The smaller LS radius enables more field lines cross the LS, and therefore leads to an increase in $\Psi_{\rm last}$. But this is obviously a minor effect compared with the decrease in $\Psi_{\rm last}$ arsing from more compact field lines in curved spacetime.

\subsection{Luminosity}

The Poynting luminosity of the pulsar is obtained by \citep{Gruz05}
\begin{eqnarray}
    L&=& \Omega \int_{0}^{\Psi_{\rm last}} I(\Psi) d\Psi\ .
\end{eqnarray}
In flat spacetime, we find the open magnetic flux and
and the corresponding luminosity are
\begin{eqnarray}
    \label{eq:Lflat}
    \Psi_{\rm last}&=&1.272 \  B_1 r_{\rm NS}^3\Omega ,\nonumber\\
    L_{\rm flat}&=&0.992 \ B_1^2 r_{\rm NS}^6\Omega^4.
\end{eqnarray}
These numbers agree well with what found in previous works \citep[e.g.][]{Gruz05,Timo06,McKinney2006,Komissarov2006,Spitkovsky2006,Kalapotharakos2009}.
For a given $\Omega$, the luminosity is completely determined by the poloidal current $I(\Psi)$ and the magnetic flux of open field lines $\Psi_{\rm last}$. We now examine how the GR effect changes each of them.

\begin{figure}
\includegraphics[width=\columnwidth]{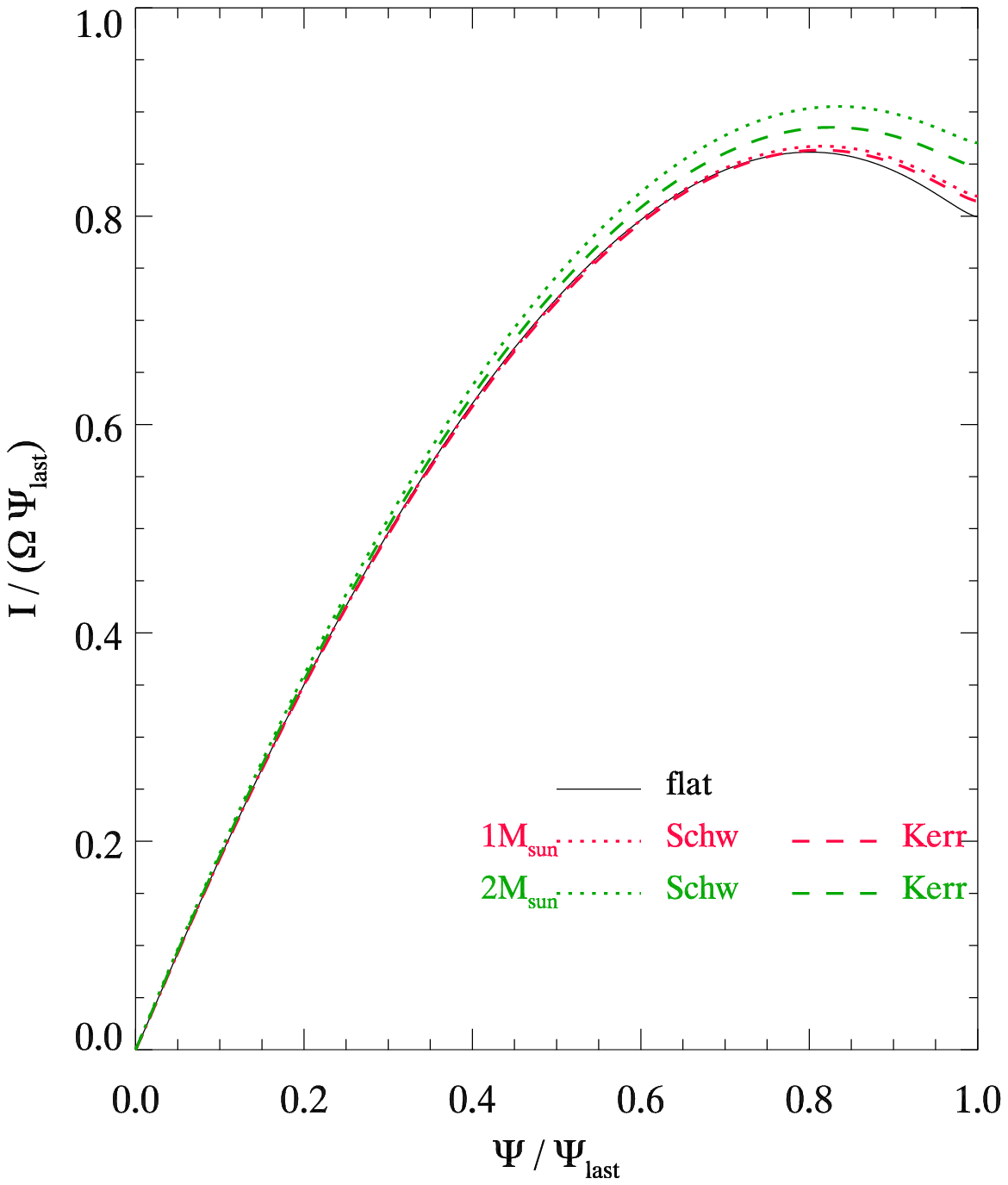}
\caption{\label{fig:NSI}
Rescaled poloidal currents $I/(\Omega \Psi_{\rm last})$ in  flat/Schwarzschild/Kerr spacetime,
are plotted in solid/dotted/dashed lines, respectively.
The red lines represent Pulsar 2, $(\Omega r_{\rm NS},M)=(0.2,1M_\odot)$,
and the green lines represent Pulsar 3, $(\Omega r_{\rm NS},M)=(0.2,2M_\odot)$.
}
\end{figure}

In Figure \ref{fig:NSI}, we compare the  poloidal currents $I(\Psi)$ of the flat/Schwarzschild/Kerr spacetime solutions for the fiducial pulsars. It is of no surprise to find that the normalized quantity $I/(\Omega \Psi_{\rm last})$ is not sensitive to the properties of the central NS or the underlying spacetime metric assumed,
since the GR corrections $(\alpha-1)$ and $(\beta-1)$ at the LS  are  small quantities of
$ \mathcal O (M/r_{\rm LS})$ and $ \mathcal O (Mr_{\rm NS}^2/r_{\rm LS}^3)$, respectively.

\begin{figure}
    \includegraphics[width=\columnwidth]{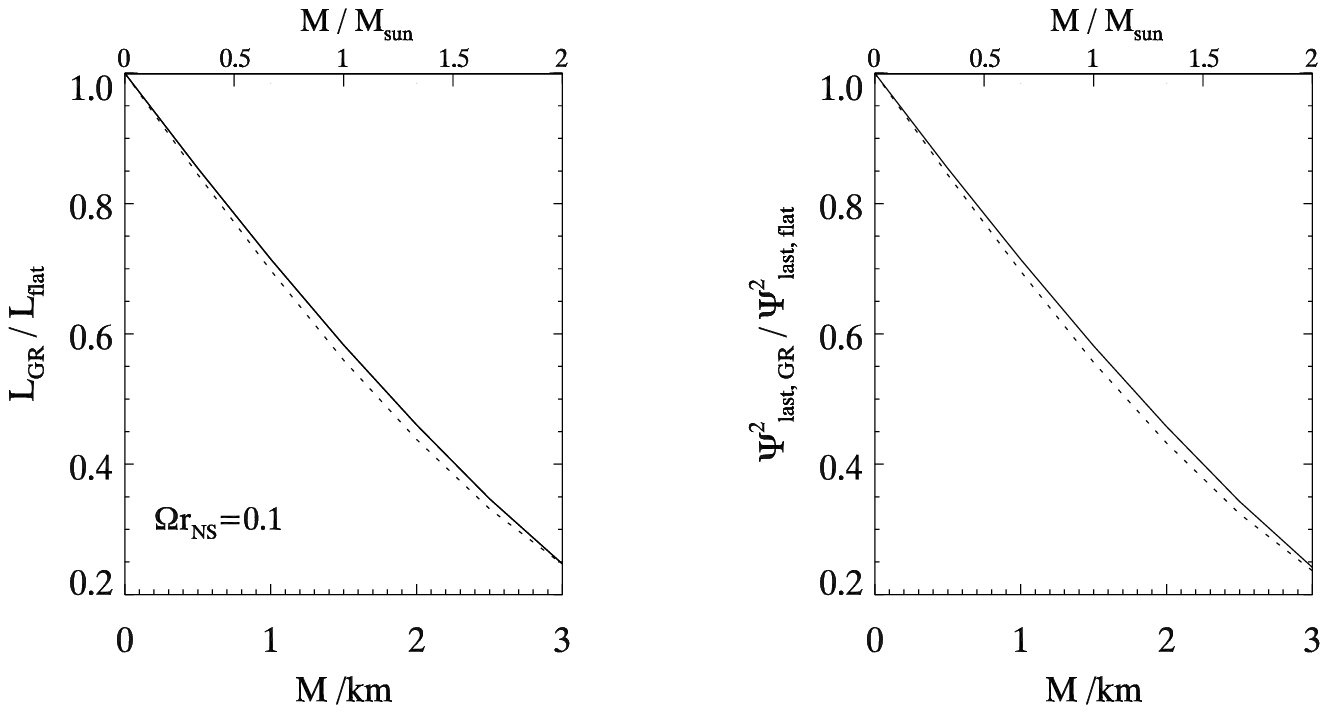}
    \includegraphics[width=\columnwidth]{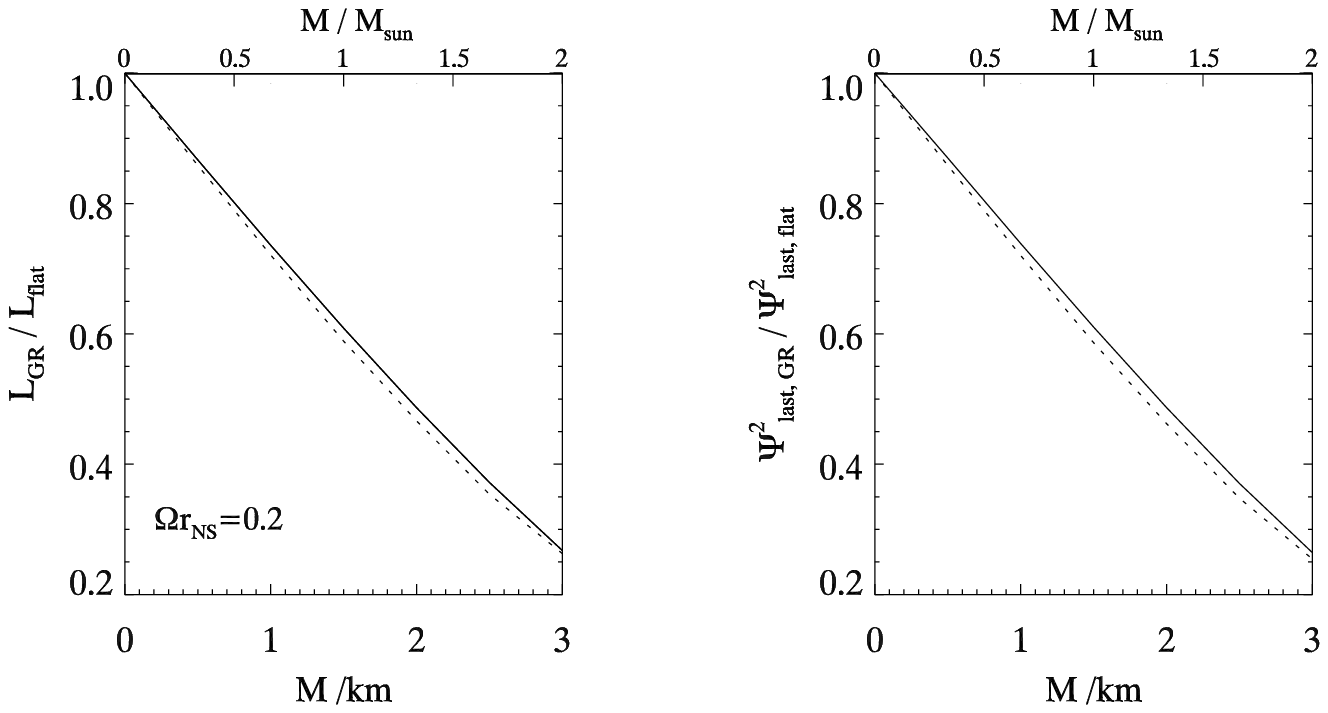}
\caption{\label{fig:LPsi2}
GR corrections to the luminosity $L_{\rm GR}/L_{\rm flat}$ and field configuration
$\Psi^2_{\rm last,GR}/\Psi^2_{\rm last,flat}$.
The upper/bottom panels show results of pulsars with $\Omega r_{\rm NS}=0.1/0.2$.
The solid/dotted lines represents results in Kerr/Shwarzschild spacetime.
 }
\end{figure}

We summarize the GR corrections to the field configuration and luminosity for pulsars with $\Omega r_{\rm NS}=0.1$ and $\Omega r_{\rm NS}=0.2$  in Figure \ref{fig:LPsi2}. We find the open magnetic flux and the Poynting luminosity decrease in the same way in curved spacetime with  increasing  NS mass $M$, where the decrease is dominated by the increased curvature $(\alpha < 1)$, while the frame-dragging effect ($\beta$ > 1) only contributes a minor part. From Figure \ref{fig:LPsi2}, we also see that the ratio $\Psi^2_{\rm last, GR}/\Psi^2_{\rm last, flat}$
and therefore $L_{\rm GR}/L_{\rm flat}$ have little dependence on the magnitude of angular velocity $\Omega$.
Base on this observation, we numerically fit the GR effect induced Poynting luminosity decrease
as a function of the central NS mass alone as follows,\footnote{The GR corrections depend on what quantity to fix doing the comparison to flat spacetime. As shown above, the GR effect gives rise to a suppression in the open magnetic flux and the Poynting luminosity if we fix the magnetic flux on the NS surface. Instead if we fix the open magnetic flux, we would find very little GR correction to the luminosity.  In some previous simulation works \citep{Ruiz14,Philippov2015,Petri16,Carrasco2018}, the luminosity comparison was done by fixing the asymptotic magnetic moment and they found that the GR effect leads to an enhancement in the open magnetic flux and the luminosity \cite[see also][for related discussions]{Gralla16}}.
\begin{eqnarray}
\label{eq:Lcorrect}
    \frac{L_{\rm Kerr}(M)}{L_{\rm flat}}&\simeq& 1 - 0.279\ \frac{M}{\rm km} + 0.011 \left( \frac{M}{\rm km} \right)^2   \nonumber\\
    &\simeq& 1 - 0.419\ \frac{M}{M_\odot} + 0.025 \left( \frac{M}{M_\odot} \right)^2 \ .
\end{eqnarray}
Due to its independence of $\Omega$, we expect this relation can be extended to slow-rotation pulsars.

Assuming the same magnetic field on the NS surface [Equation (\ref{eq:bc})],  we find a $\sim 73$ percent decease in
the Poynting luminosity arising from GR correction $L_{\rm Kerr}(2M_\odot)/L_{\rm flat} = 0.27$ for Pulsar 3.
In the reverse direction, we may estimate the magnetic field strength on the NS surface from the observed luminosity $L_{\rm obs}$,
using the relation $L_{\rm obs}\propto \Psi_{\rm last}^2 \propto  B_1^2$ [Equation (\ref{eq:Lflat}) and (\ref{eq:Lcorrect})]. We would obtain two different results
$B_{1,\rm Kerr}$ versus $B_{1, \rm flat}$ depending on whether the GR effect is taken account of or not.
For Pulsar 3, it  it is clear that the two estimates are related by $B_{1, \rm Kerr}(2M_\odot)/B_{1, \rm flat}=\sqrt{1/0.27}=1.93$,
i.e., a $93\%$ increase in the surface field estimate if the GR effect is taken account of.
To compare our numerical results with previous works, we consider another pulsar with mass of $1.67M_\odot$ (or $M=2.5{\rm km}$),  and we find $B_{1, \rm Kerr}(1.67M_\odot)/B_{1, \rm flat}= \sqrt{1/0.37} = 1.64$, i.e., an  $ 64\%$ increase in the $B_1$ estimate. This result is highly consistent with the recent simulation+analytic study for a slow-rotation pulsar with same NS mass but much smaller angular velocity $\Omega r_{\rm NS} = 0.02$ \citep{Gralla16}. This agreement again verifies that the luminosity suppression [Equation (\ref{eq:Lcorrect})] arising from the GR corrections is insensitive to the NS angular velocity $\Omega$.

\subsection{Current on the Polar Cap}
As discussed in the introduction, spacelike current in the force-free magnetospshere is
an indicator of pair production. Now we numerically pin down these regions.
Accurate to the leading order of $\Omega r_{\rm NS}$, the electric current 4-vector is expressed as \citep[e.g.][]{Philippov2015,Gralla16}
\begin{eqnarray}
    \label{eq:j2}
    J^t&=& \frac{2\Omega }{\alpha^2\beta r^2} \left[ (r-3M) \partial_r\Psi + \cot\theta \partial_\theta \Psi \right]\ ,  \nonumber\\
    J^r&=& - \frac{1}{r^2\sin\theta} \partial_\theta I(\Psi)\ , \nonumber\\
    J^\theta&=& \frac{1}{r^2\sin\theta} \partial_r I(\Psi)\ , \ J^\phi=0\ ,  \nonumber\\
    J^2&=& -\alpha^2 J^tJ^t + \alpha^{-2} J^rJ^r + r^2 J^\theta J^\theta\ .
\end{eqnarray}

\begin{figure*}
    \includegraphics[scale=0.38]{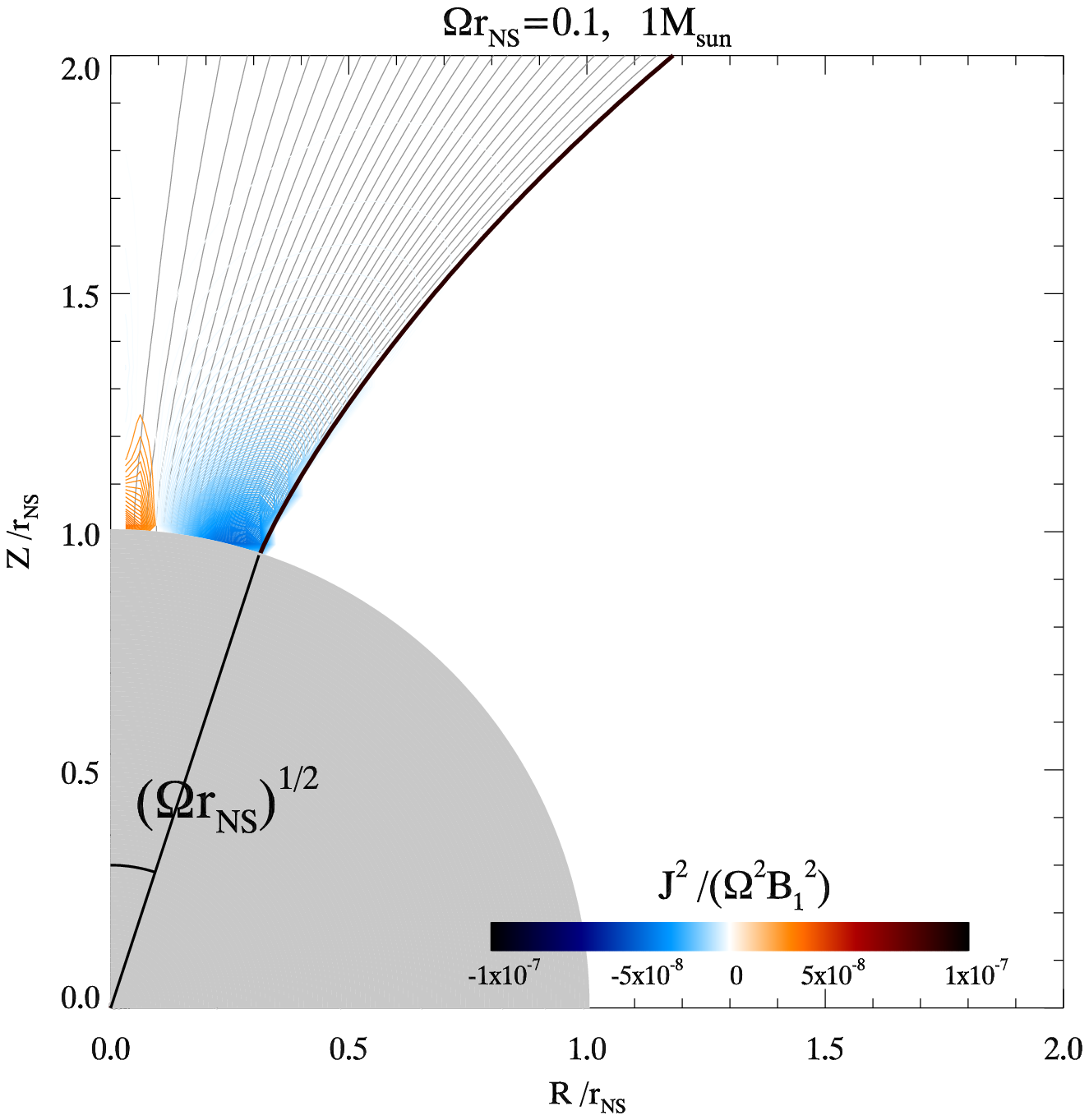}
    \includegraphics[scale=0.38]{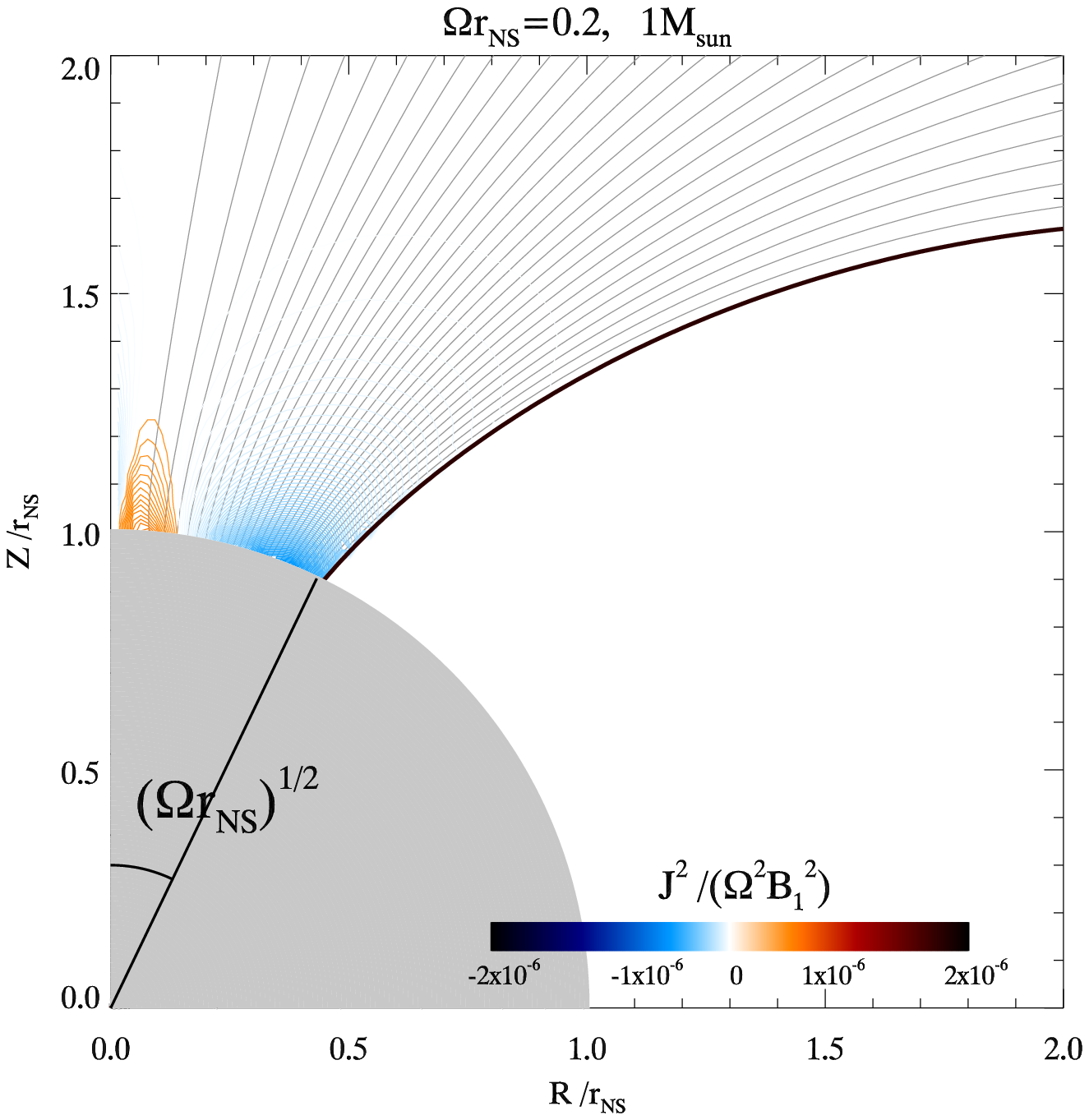}
    \includegraphics[scale=0.38]{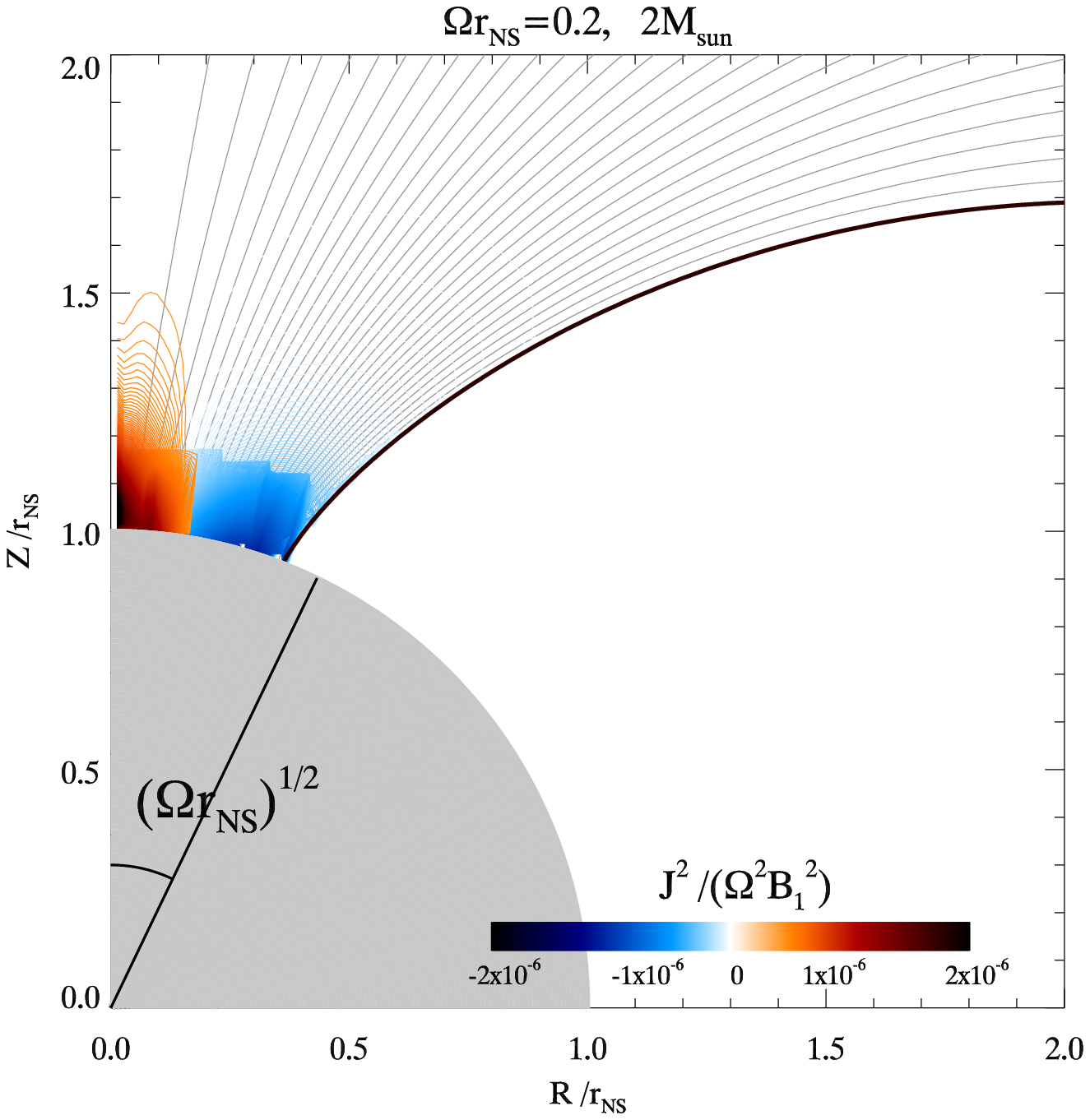}
\caption{\label{fig:J2}
Current $J^2$ contours for Kerr solutions of the fiducial pulsars, where the red/blue contours represent
the spacelike/timelike  current. The open field lines of dipolar field are also shown in the background,
where the last open field line (thick black line) roots on the NS surface at $\theta_{\rm cap} \sim (\Omega r_{\rm NS})^{1/2}$.
 }
\end{figure*}

For either flat or Schwarzschild solution, we find no spacelike current for the dipolar magnetosphere,
which is consistent with previous studies \citep{Philippov2015,Gralla16}, indicating that
the spacelike current in the dipolar magnetosphere is completely generated by the frame-dragging effect.
We plot the $J^2$ contours along with the open lines for Kerr solutions of the three fiducial pulsars in Figure \ref{fig:J2}, where the red contours represent the spacelike current $J_+^2$ ($J^2>0$), and the blue contours represent the timelike current $J_-^2$ ($J^2<0$).

\begin{figure}
    \includegraphics[width=\columnwidth]{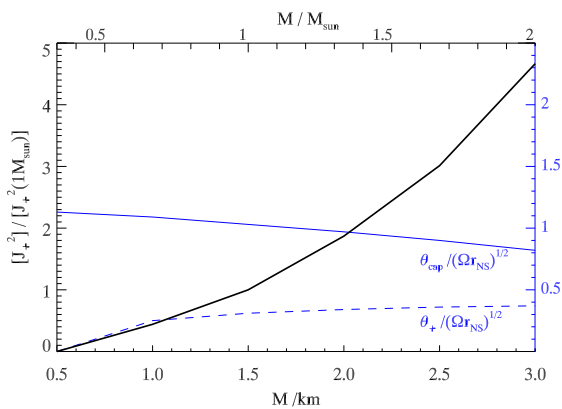}
\caption{\label{fig:aveJp2}
{\em Solid black:} the relation between averaged amplitude of spacelike current
$\left[J_+^2 \right]/\left[J_+^2(1M_\odot)\right]$
and the neutron star mass $M$ in aligned dipole magnetosphere.
{\em Solid blue:} the size of non-zero current region $\theta_{\rm cap}$ on the NS surface as function of $M$ and $\Omega$. {\em Dashed blue:} the size of spacelike current region $\theta_+$.
 }
\end{figure}

Now we proceed to examine how the current distribution depends on the angular velocity $\Omega$ and the NS mass $M$ via the curvature effect and the frame-dragging effect. For this purpose, we define two angles, $\theta_{\rm cap}$ where the last-open-field line roots on the NS surface, and  $\theta_+$ which is the boundary between $J^2_+$ and $J^2_-$ on the NS surface, and an averaged quantity
\begin{eqnarray}
    \left[J_+^2\right] &=& \frac{\int\int^{J^2>0} J^2(R,Z)\ {\rm d}R {\rm d}Z }{ \int\int^{J^2>0} {\rm d}R {\rm d}Z } \ , \nonumber
\end{eqnarray}
which quantifies the intensity of spacelike current.
In Figure \ref{fig:aveJp2}, we show the numerical results of these three quantities as functions of $\Omega$ and $M$.

The dependence on $\Omega$ is easy to understand.  It is straightforward to see that
$\theta_{\rm cap}\sim (\Omega r_{\rm NS})^{1/2}$ from the open magnetic flux $\Psi_{\rm last}$ and the boundary condition Equation (\ref{eq:bc}), and $J^2\propto \Omega^2 B_1^2$ from Equation (\ref{eq:j2}). These simple scaling relations explain what shown in Figure \ref{fig:aveJp2}: $\theta_+ \lesssim \theta_{\rm cap} \sim (\Omega r_{\rm NS})^{1/2}$ and $\left[J_+^2 \right]/\left[J_+^2(1M_\odot)\right]$ has no dependence on $\Omega$.

There is no such simple scaling relation for  the dependence on $M$, but its qualitative behavior is also easy to understand.  As shown in Figure \ref{fig:NS02} and \ref{fig:LPsi2}, the field lines become more compact and the open magnetic flux $\Psi_{\rm last}$ decreases with increasing mass $M$ due to the curvature effect, which lead to an increase in the magnitude of $J^2$ (both $J_+^2$ and $J_-^2$) and a slight decrease in $\theta_{\rm cap}$, respectively (see the latter two panels of Figure \ref{fig:J2}).  From Equation (\ref{eq:j2}), we see that the charge density $J^t$ decreases with increasing NS mass $M$ due to the frame-dragging effect ($\beta > 1$). As a result, $J^2$ becomes more positive and therefore $\theta_+$ increases with increasing $M$.  Specifically, spacelike current exists in region near the NS surface with $\theta \leq \theta_+= 0.29\ \theta_{\rm cap}$ for Pulsar 1 and Pulsar 2 with $M=1M_\odot$, and the spacelike-current region expands to $\theta \leq \theta_+ = 0.48\  \theta_{\rm cap}$ for Pulsar 3 with $M=2M_\odot$.

\section{GR Effect in Pulsar Magnetosphere Structure: Aligned Multipoles}
\label{sec:GRm}

In this section, we investigate the GR corrections in the pulsar magnetosphere with multipolar boundary condition.
As a simple example, we consider the superposition of a dipole field and an octupole field,
\begin{equation}
    \Psi(r_{\rm NS}, \theta) = B_1 r_{\rm NS}^2 \left[ \sin^2\theta + a_1\ (1-5\cos^2\theta) \sin^2\theta \right] \ .
\end{equation}

\subsection{Luminosity}

\begin{figure*}
    \includegraphics[scale=0.45]{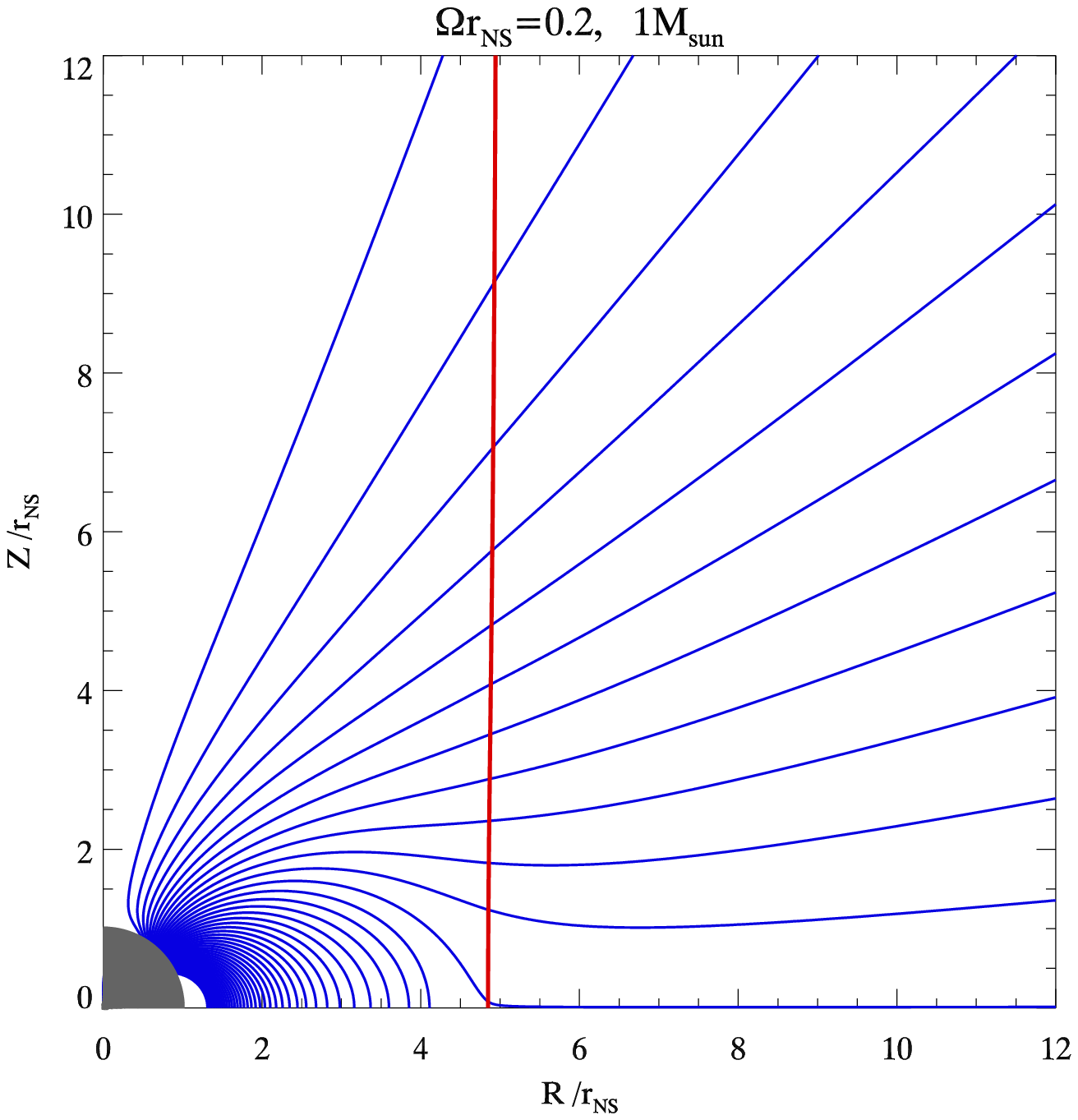}
    \includegraphics[scale=0.45]{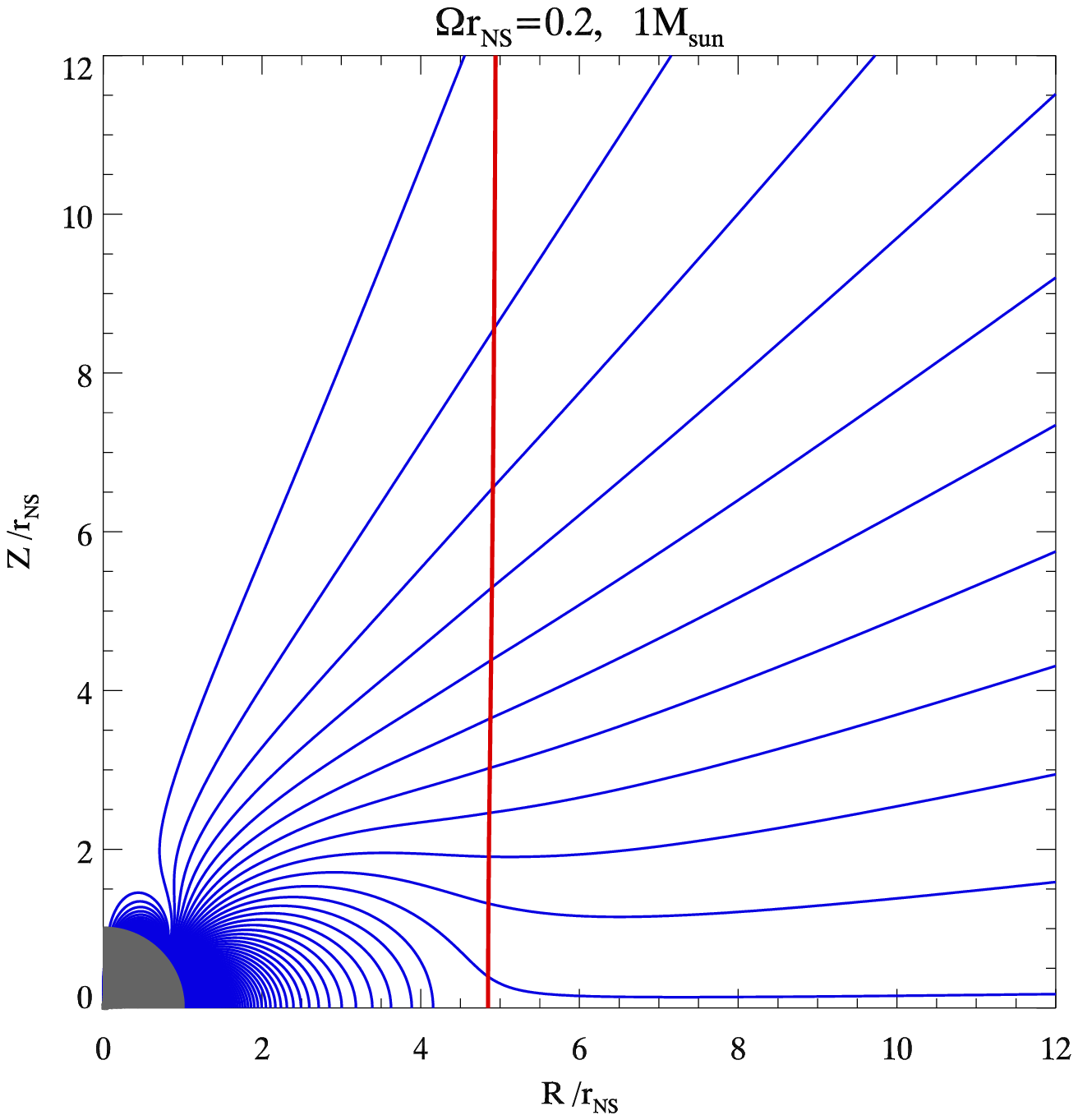}
\caption{\label{fig:New_oc+}
Field-line configuration of Pulsar 2 with different octupole components $a_1$ in the boundaries.
{\em Left}: $a_1=1/3$; {\em Right}: $a_1=1$.
The red lines represent the bent LS.
 }
\end{figure*}

\begin{figure*}
    \includegraphics[scale=0.38]{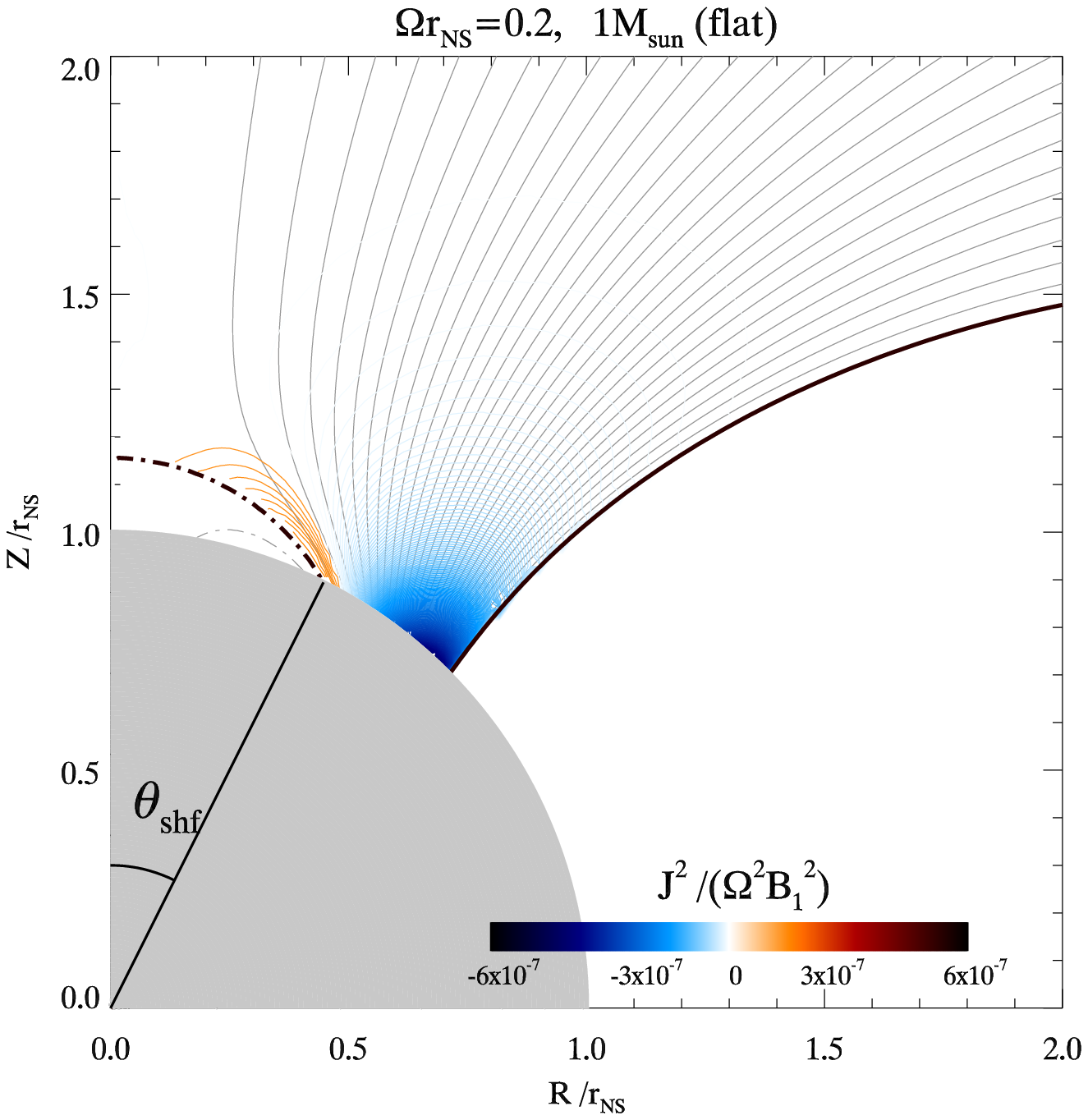}
    \includegraphics[scale=0.38]{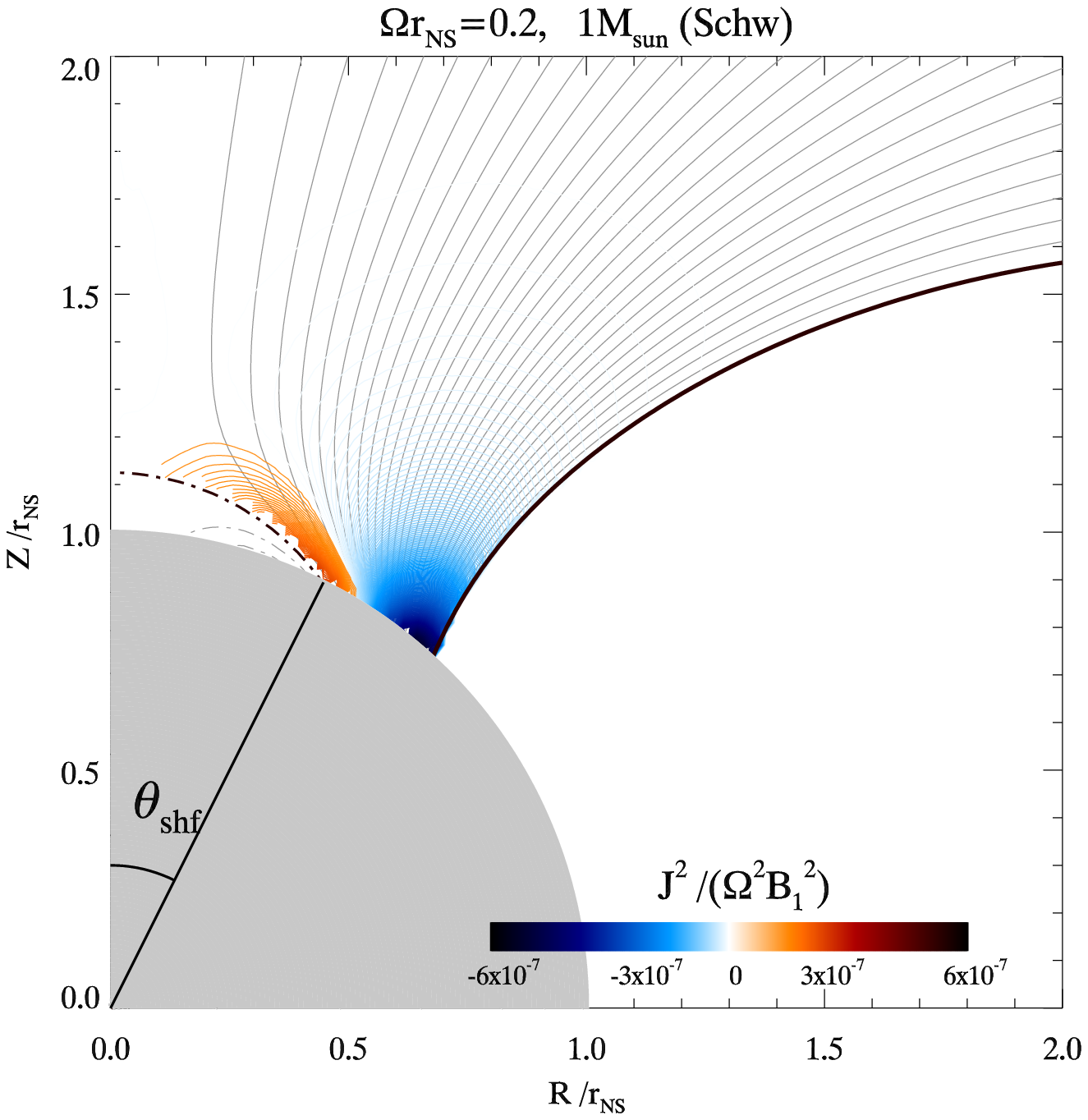}
    \includegraphics[scale=0.38]{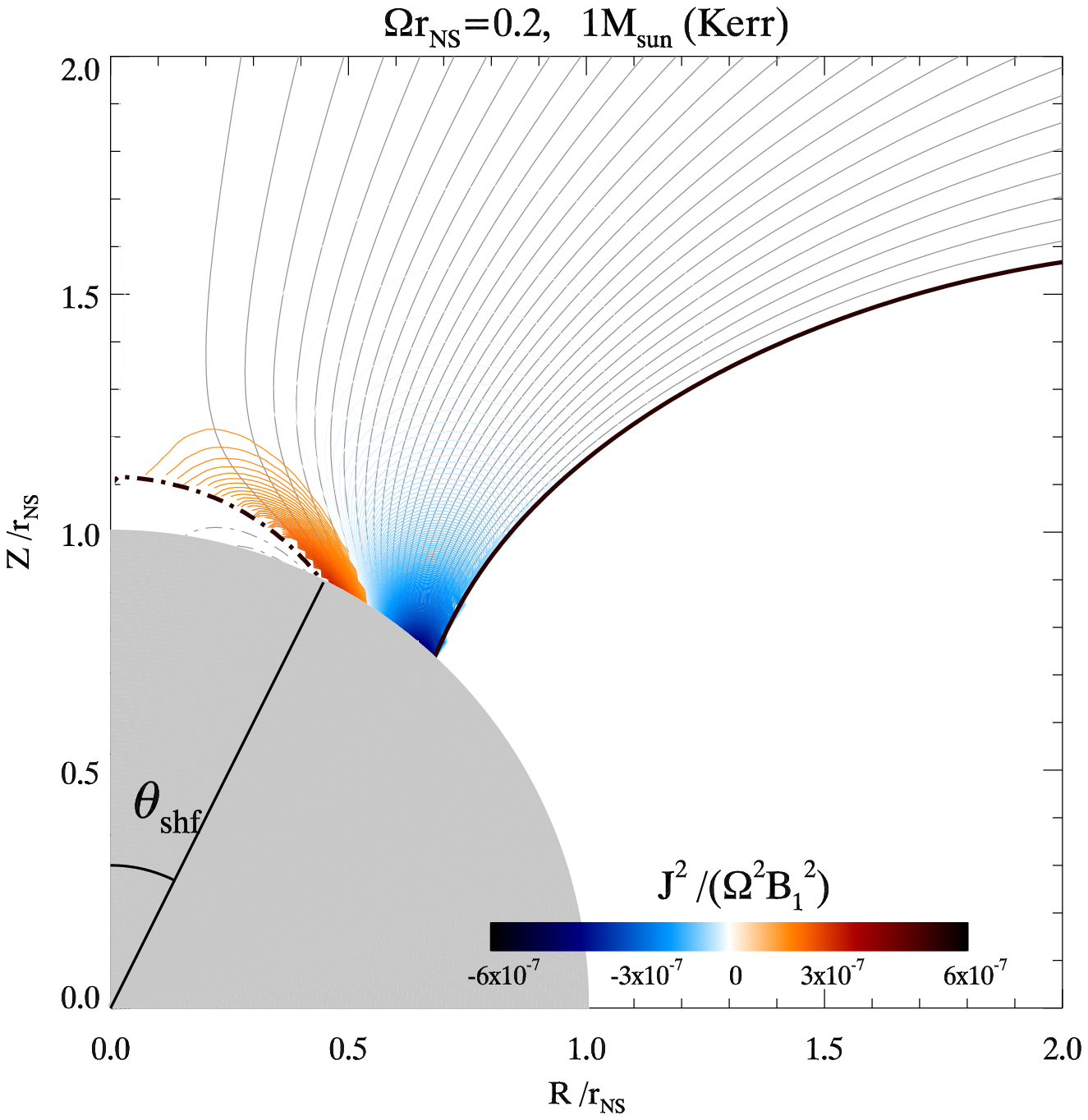}
\caption{\label{fig:oc+1o3}
Contours of current $J^2$ in polar cap of Pulsar 2 with multipolar boundary condition $a_1=1/3$.
{\em Left}: in flat spacetime; {\em Middle}: in Schwarzschild spacetime; {\em Right}: in Kerr spacetime.
The red/blue contours represent the space-like/time-like current.
The open-lines of dipolar field are shown solid lines ($\Psi>0$) and dotted-dashed lines ($\Psi<0$).
The field-line with $\Psi=0$ is shown in the thick dotted-dashed line.
The last closed-line $\Psi_{\rm last}$ is shown in thick black line.
The polar cap region $J^2\ne0$ shifts away from the pole by $\theta_{\rm shf}$.  }
\end{figure*}

\begin{figure*}
    \includegraphics[scale=0.38]{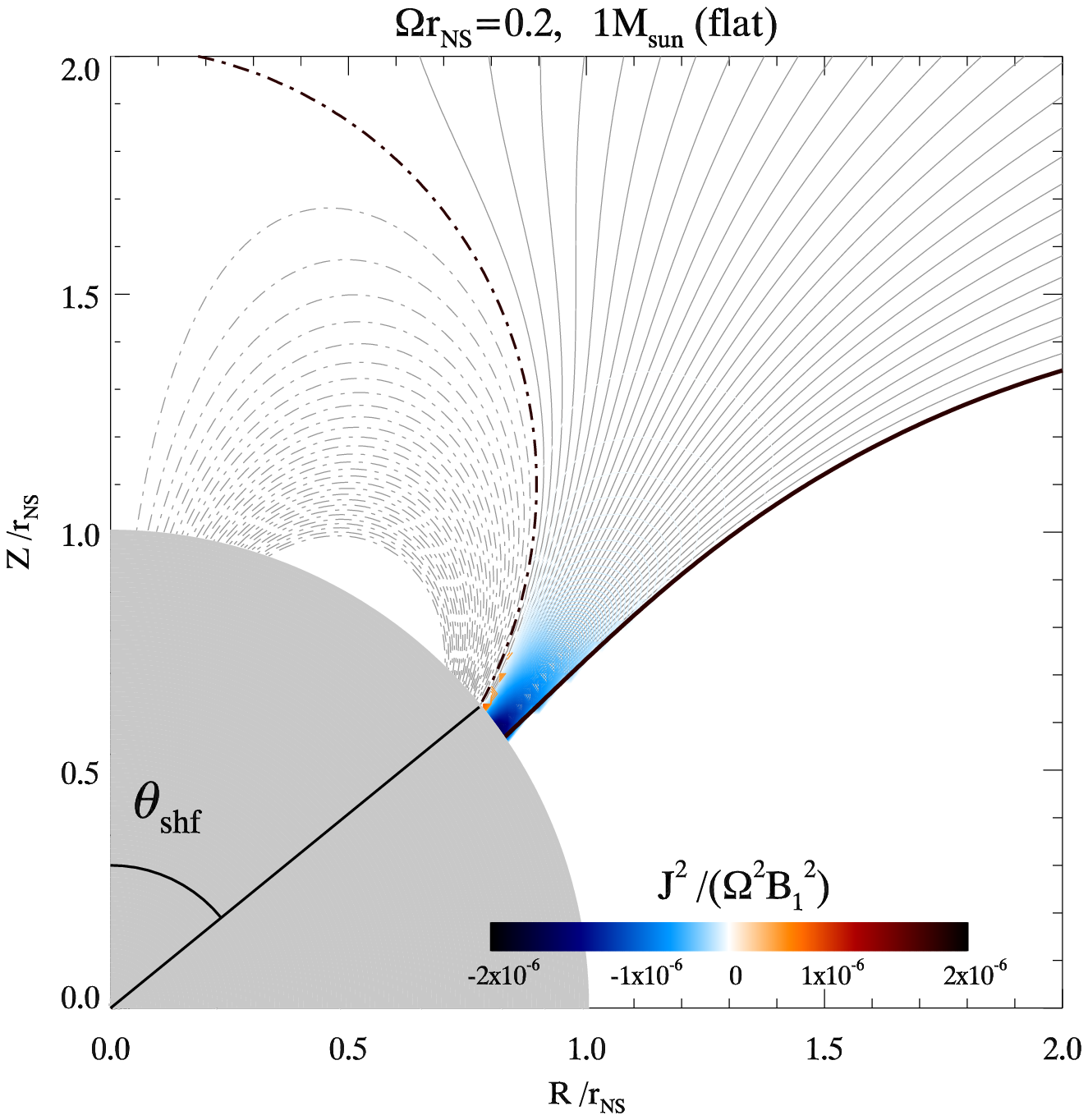}
    \includegraphics[scale=0.38]{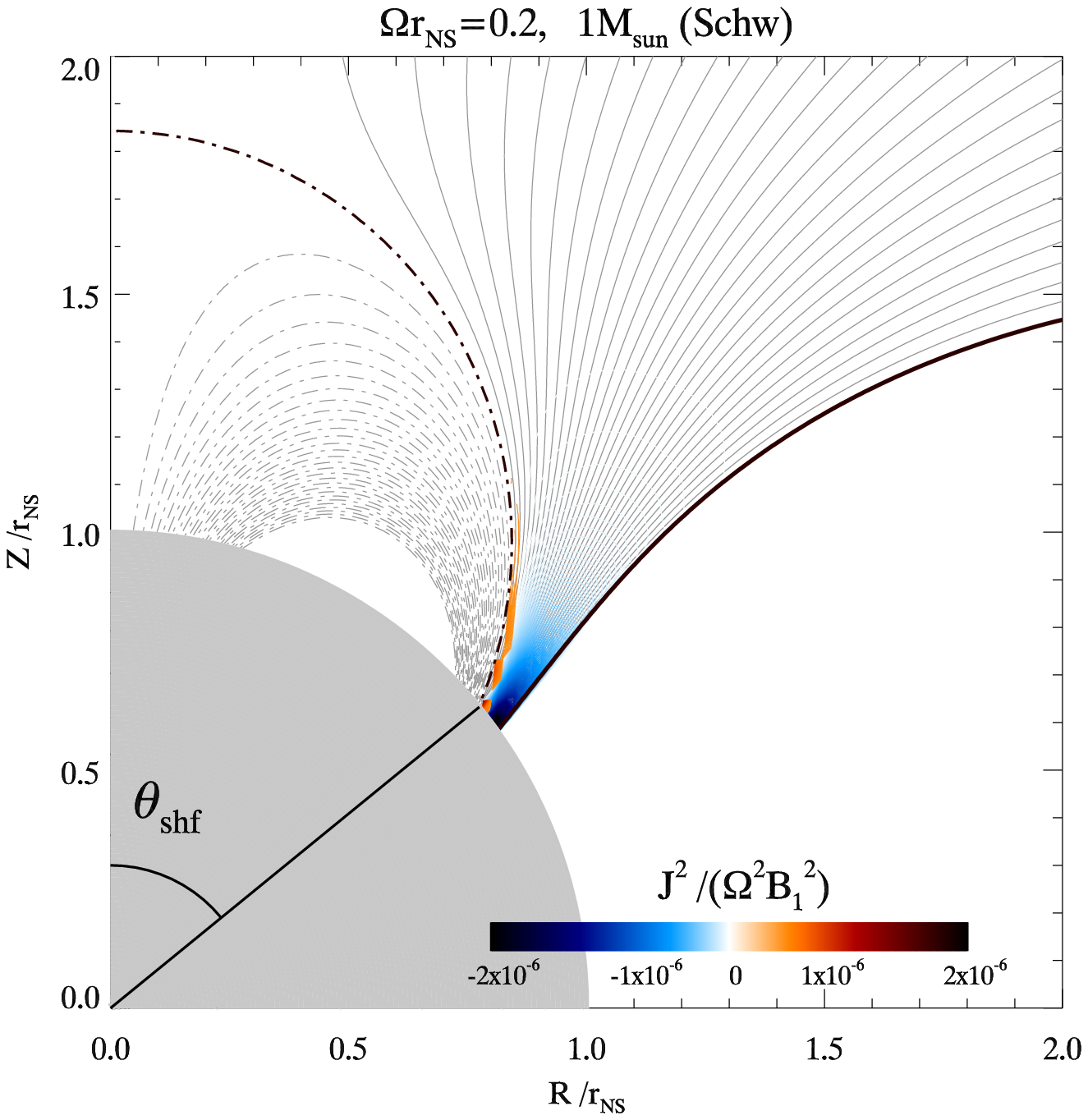}
    \includegraphics[scale=0.38]{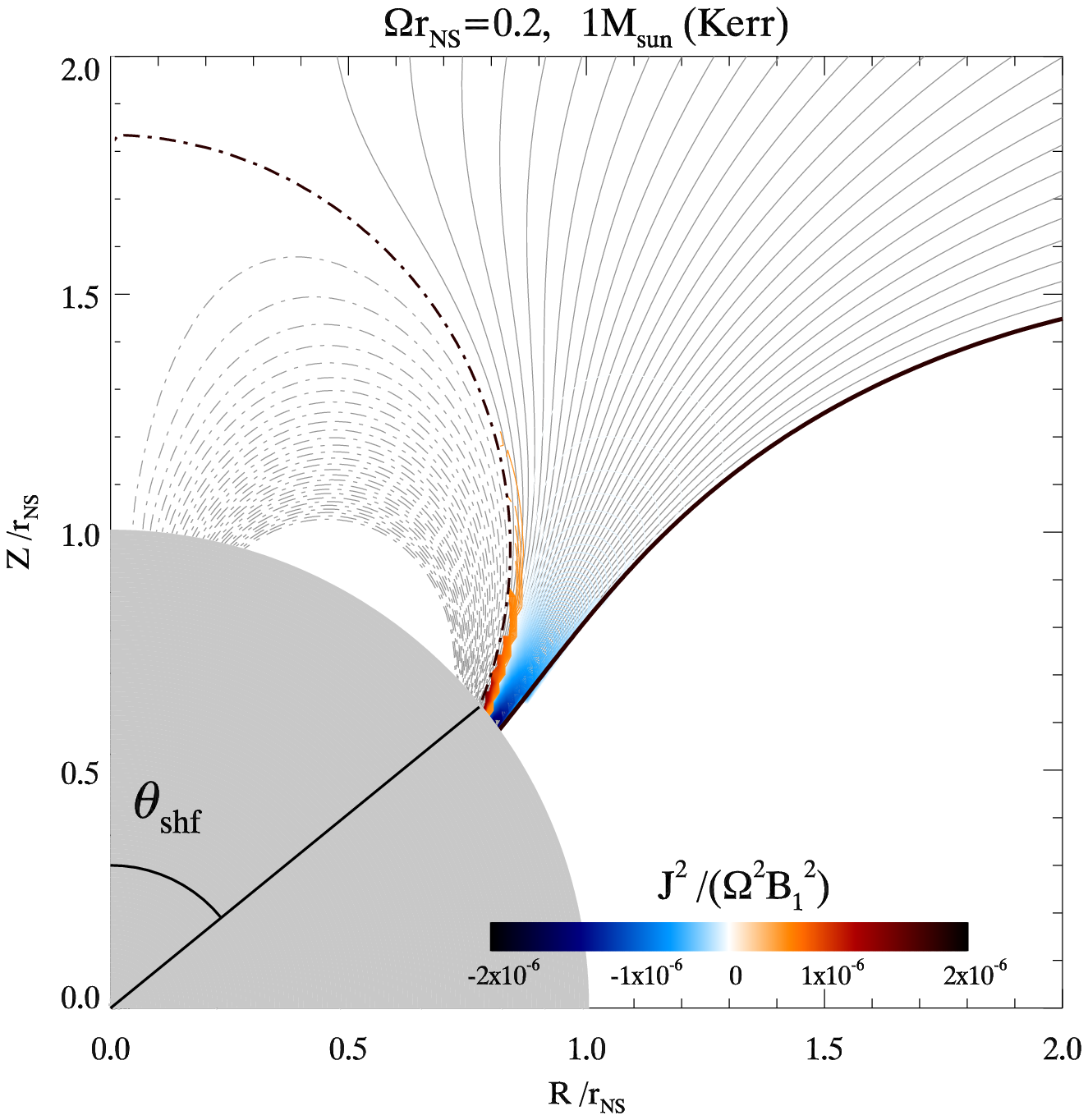}
\caption{\label{fig:oc+1}
Similar to Figure \ref{fig:oc+1o3}, but with $a_1=1$.
 }
\end{figure*}

The field configuration is sensitive to the octupole component $a_1$. Specifically, for $a_1>1/4$, there are two magnetic poles where $\Psi =0$:
other than the usual  $\theta =0$ one, another shifts away from the former by
\begin{eqnarray}
    \theta_{\rm shf}&=& \arctan \sqrt{ 4 - 5 (1+a_1)^{-1} }\ .
\end{eqnarray}
In Figure \ref{fig:New_oc+}, we show the field line configurations for Pulsar 2 with octupole component $a_1=1/3$ and $a_1=1$, where the octupole component only changes the field configuration close to the NS, since the high-order component decreases faster with increasing $r$ than the dipole component. Therefore we expect the octupole component does not affect the Poynting luminosity which is sensitive to the open magnetic flux $\Psi_{\rm last}$.  Numerically we find
\begin{equation}
\begin{aligned}
    \Psi_{\rm flat, a_1=\frac{1}{3}} &= 1.299 \ B_1 r_{\rm NS}^3 \Omega \ , &L_{\rm flat, a_1=\frac{1}{3}} &= 1.042 \ B_1^2 r_{\rm NS}^6 \Omega^4 ,\\
    \Psi_{\rm flat, a_1=1} &=  1.355 \ B_1 r_{\rm NS}^3 \Omega \ ,  &L_{\rm flat, a_1=1} &= 1.143 \ B_1^2 r_{\rm NS}^6 \Omega^4 ,
\end{aligned}
\end{equation}
and the fitting formula (\ref{eq:Lcorrect})
quantifies the GR effect induced luminosity suppression for millisecond pulsars accurate to  $\sim 5\%$
as $a_1$ varies from $-1$ to $1$.

\subsection{Multipolar Annuli: $a_1>1/4$}

The octupole  component makes little difference to the Poynting luminosity which is sensitive to the magnetic field around the LS, but makes a big difference in generating spacelike current which is sensitive to the field near the NS.
In Figure \ref{fig:oc+1o3} and Figure \ref{fig:oc+1},
we plot the field lines with $\Psi\le\Psi_{\rm last}$ for Pulsar 2 with octupole component $a_1=1/3$ and $a_1=1$, respectively,
where the dotted-dashed lines denote field lines with negative magnetic flux ($\Psi<0$),
the solid lines denote  field lines with positive magnetic flux ($\Psi>0$),
the thick dotted-dashed line denote field line with $\Psi=0$,
and the last closed field-line $\Psi_{\rm last}$ is plotted as the thick black line.

From Figure \ref{fig:oc+1o3} and Figure \ref{fig:oc+1}, we see the current distribution shows some
new features due to the presence of the octupole component. First, the magnetic polar cap characterized by non-zero current $J^\mu$ shifts away from the  pole of the NS; as a result, the polar cap turns to a narrow annulus around the star's pole.  Second, the size of the magnetic polar cap $\theta_{\rm cap}$ no longer has a simple scaling relation with $\Omega$ due to the extra dependence on the octupole component $a_1$.
Third, spacelike current shows up in both flat and Schwarzschild spacetime solutions, therefore the frame-dragging effect is no longer the necessary condition for generating spacelike current, as pointed out by \citet{Gralla16}.

Other than these differences, our analysis
about the current distribution of the dipolar magnetosphere dependence on the curvature effect and the frame-dragging effect in Section \ref{sec:GR} also largely applies here. Specifically, the curvature effect compactifies the magnetic field lines around the NS and decreases the open magnetic flux, consequently increases the magnitude of $J^2$ (both $J_+^2$ and $J_-^2$) and decreases the magnetic polar cap size $\theta_{\rm cap}$. The frame-dragging effect decreases the charge density, therefore increases both the size of the spacelike-current region $\theta_+$ and the magnitude of $J_+^2$.

Here we only consider the structure of axisymmetric magnetosphere with aligned dipole and octupole fields.
The annuli structure of spacelike current in presence of octupole field and its sensitive dependence on the octupole
component indicates rich patchy patterns of pair-production regions for pulsars with more complicate magnetic field
configurations, e.g. inclined magnetic field with several different moments of the same order of magnitude \citep{Gralla2017}.
The patchy patterns of the pair-production regions are desired by the patchy beam models for radio pulsars,
e.g. the fan beam model proposed by \cite{Wang14}, where a large amount of electrons and positrons are produced within only a few separated magnetic flux tubes, and these particles produce coherent emission as they move along the field lines, while other inactive flux tubes fail to produce emission.

\section{Summary and Discussion}
\label{sec:summary}

The structure of force-free NS magnetospheres is governed by a second-order differential equation, GS equation, of the toroidal component of the vector potential $\Psi$, which degrades to first order at the LS. The global solution to the GS equation in flat spacetime has been investigated in many previous works. The basic strategy to numerically solve the GS equation is to adjust the poloidal current $I(\Psi)$  and ensure the field lines to smoothly cross the LS, which is a straight line in flat spacetime. But implementing this algorithm becomes complicate if the LS is not straight, e.g., in Schwarzschild or Kerr spacetime, because there is no simple computation grid  adapted to the bent LS. To avoid this grid-LS mismatch complication, we  introduced a new coordinate frame designed such that the LS in it is a straight line. Then we could numerically solve the curved-spacetime GS equation using the familiar cross-straight-LS algorithm.

As an application, we numerically solved the curved-spacetime GS equations for millisecond pulsars with aligned dipole magnetic fields and systematically examined the GR corrections to the structure of pulsar magnetospheres, including the Poynting luminosity and the current distribution, depending on the angular velocity $\Omega$ and the NS mass $M$.
In our investigation, we split  the GR effect into two parts, curvature and frame-dragging.
With the magnetic flux on the NS surface fixed,  we found that the curvature effect compatifies the field lines near the NS, therefore gives rise to a suppression in the open magnetic flux  and a similar suppression in the Poynting luminosity.
Compared to flat spacetime, the Poynting luminosity is suppressed by $39\%$ for $1M_\odot$ NS and $73\%$ for $2M_\odot$ NS, which are independent of the NS angular velocity $\Omega$.
The luminosity reduction induced by GR effects may also have interesting implications for the recent low luminosity Gamma Ray Burst (GRB) associated with Gravitational Wave 170817 event \citep{Tong18}. 

We found that the frame-dragging effect contributes a minor part in shaping the magnetospheres even for millisecond pulsars, but plays a role in generating spacelike current. We also numerically examined and
qualitatively explained the spacelike current dependence on $\Omega$ and $M$ via the curvature effect and the frame-dragging effect. We found the polar cap size $\theta_{\rm cap}$, the spacelike current size $\theta_{+}$
and the averaged amplitude of spacelike current $\left[J^2_+ \right]$ all have simple scaling relations
with $\Omega$. While their dependence on $M$ is more subtle, and mainly comes from the curvature effect induced
more compact magnetic field configuration and frame-dragging effect induced low charge density.

We did similar analysis for pulsars with both dipole field and octupole field.
The presence of octupole field greatly enriches the structure of spacelike-current regions, where pair production
likely takes place. In the axisymmetric case, we find the spacelike current is generated on narrow annuli enclosing the poles of the NS, with their locations and sizes sensitively depending the intensity of the octupole component. For realistic pulsars with more complicate magnetic field, e.g., inclined magnetic field with several multipole moments of the same order of magnitude, we expect rich patchy patterns of pair production regions, which are desired by the patchy beam models for radio pulsars.

\section*{Acknowledgements}
L.H. thanks the support by the National
Natural Science Foundation of China (grants 11203055, 11590784 and 11773054),
and Key Research Program of Frontier Sciences, CAS (grant No. QYZDJ-SSW-SLH057).
Z.P. is  supported by the Dissertation Year Fellowship of UC Davis.
C.Y. is grateful for the support by the National Natural Science Foundation of China (grants 11373064, 11521303 and 11733010),
Yunnan Natural Science Foundation (grants 2014HB048),
and Yunnan Province (2017HC018).
This work made extensive use of the NASA Astrophysics Data System and
of the {\tt astro-ph} preprint archive at {\tt arXiv.org}.


%\bibliographystyle{mnras}

\appendix

\section{new coordinates $(\tilde{r}, \tilde{\theta})$}
\label{sec:appA}

We adopt a new reference frame $(\tilde{r}, \tilde{\theta})$,
to include both the factors of curvature $\alpha$ and frame-dragging $\beta$ in the coordinate radius $\tilde{r}$
\begin{eqnarray}
    \tilde{r}&=& \frac{r}{\alpha\beta}\ =\ r \cdot\frac{\left( 1 - 2\hat{I}_{\rm NS}/r^3 \right)}{\sqrt{1 - 2M/r}}\ .
\end{eqnarray}
The first- and second- order derivatives of $r$ is related to those of $\tilde{r}$
by the following expressions
\begin{eqnarray}
    \partial_r&=&  \mathcal{P}\ \partial_{\tilde{r}} \ ,  \nonumber\\
    \partial_{rr}&=& \mathcal{P}^2\ \partial_{\tilde{r}\tilde{r}} + \mathcal{Q}\ \partial_{\tilde{r}} \ .  \nonumber\\
    {\rm where}\ \mathcal{P}&=& \frac{1}{\alpha} - \frac{M}{\alpha^3r} + \frac{4\hat{I}_{\rm NS}}{\alpha r^3} + \frac{2\hat{I}_{\rm NS}M}{ \alpha^3r^4}\ , \nonumber\\
    \mathcal{Q}&=& \frac{3M^2}{\alpha^5r^3} - \frac{12\hat{I}_{\rm NS}}{\alpha r^4} - \frac{12\hat{I}_{\rm NS}M}{\alpha^3 r^5} - \frac{6\hat{I}_{\rm NS}M^2}{ \alpha^5r^6}\ .
\end{eqnarray}
The GS equation Equation (\ref{eq:GS_1}) is then transformed  as
\begin{eqnarray}
    0&=& \mathcal{S}(\partial_{\tilde{r}\tilde{r}}, \partial_{\tilde{\theta}\tilde{\theta}}, \partial_{\tilde{r}}, \partial_{\tilde{\theta}}; \Psi) \nonumber\\
    &=& (1 - \Omega^2 \tilde{r}^2\sin^2\tilde{\theta})\ \alpha^4 \beta^2 \mathcal{P}^2\ \partial_{\tilde{r}\tilde{r}} \Psi  \nonumber\\
    &\ & + (1 - \Omega^2 \tilde{r}^2\sin^2\tilde{\theta})\ \alpha^2 \beta^2 \mathcal{Q}\ \partial_{\tilde{r}} \Psi  \nonumber\\
    &\ & + ( \alpha^{-2} - 1 - 2\Omega^2 \tilde{r}^2\sin^2\tilde{\theta})\ \alpha^3\beta \mathcal{P}\ \frac{\partial_{\tilde{r}} \Psi}{\tilde{r} }  \nonumber\\
    &\ & + (1 - \Omega^2 \tilde{r}^2\sin^2\tilde{\theta})\ \frac{1}{\tilde{r}^2} \partial_{\tilde{\theta}\tilde{\theta}} \Psi - \frac{\cos\tilde{\theta}}{\tilde{r}^2} \partial_{\tilde{\theta}} \Psi                    \nonumber\\
    &\ & - \Omega^2 \sin^2\tilde{\theta} \cos\tilde{\theta} \partial_{\tilde{\theta}} \Psi
     + \beta^2 \frac{II_{,\Psi}}{4\pi^2}\ .
\end{eqnarray}

\section{coefficients in GS Equation (\ref{EQ:GS_2}) }
\label{sec:appB}

We define intermediate coefficients $\mathcal{A},\mathcal{B},\mathcal{C}$
related to factors $\alpha$ and $\beta$ as
\begin{eqnarray}
    \mathcal{A}&=& \left[ \frac{3\alpha^2-1}{2\alpha}\ \beta\ +\  \frac{3\alpha^2+1}{2\alpha}\ (\beta-1) \right]^2 - 1 \ ,  \nonumber\\
    \mathcal{B}&=& (3-3\alpha^2)\ \beta - \frac{3(\alpha^2-1)^2}{4\alpha^2}\ \beta(2-\beta)\ , \\
    \mathcal{C}&=& \frac{3(\alpha^2-1)^2}{4\alpha^2}\ \beta(2-\beta)\ -\ 3(\alpha^2+1)(\beta-1)  \nonumber\\
    &\ & +\ \frac{(\alpha^{-2}-1)}{2} \left[ (3\alpha^2-1)\ \beta\ +\ (3\alpha^2+1)\ (\beta-1) \right] \ . \nonumber
\end{eqnarray}
The coefficients $\mathcal{D,E,F,G,H}$ in Equation (\ref{EQ:GS_2}) have explicitly expressions as follows
\begin{eqnarray}
    \mathcal{D}&=& \beta^{-2}\left( 1 + \mathcal{A}\ \frac{\tilde{R}^2}{\tilde{R}^2+\tilde{Z}^2} \right)\ , \nonumber\\
    \mathcal{E}&=& \beta^{-2}\left(1 + \mathcal{A}\ \frac{\tilde{Z}^2}{\tilde{R}^2+\tilde{Z}^2} \right)\ , \nonumber\\
    \mathcal{F}&=& \beta^{-2} (\mathcal{B}\ \Omega^2\tilde{R}^2 + \mathcal{C}) \frac{\tilde{R}^2}{\tilde{R}^2+\tilde{Z}^2}\ ,  \nonumber\\
    \mathcal{G}&=& \beta^{-2} (\mathcal{B}\ \Omega^2\tilde{R}^2 + \mathcal{C}) \frac{\tilde{Z}^2}{\tilde{R}^2+\tilde{Z}^2}\ ,  \nonumber\\
    \mathcal{H}&=& 2\beta^{-2} \mathcal{A}\ \frac{\tilde{R}\tilde{Z}}{\tilde{R}^2+\tilde{Z}^2}\ .
\end{eqnarray}

\section{coefficients in L'H\^opital's rule}
\label{sec:appC}

In new reference frame $(\tilde{r},\tilde{\theta})$,
the curvature factor is
\begin{eqnarray}
\label{eq:alpha}
    \alpha^2&=& 1 - \frac{2M}{\alpha\beta\tilde{r}}\ ,
\end{eqnarray}
and the frame-dragging factor is
\begin{eqnarray}
\label{eq:beta}
    \frac{1}{\beta}&=& 1 - \frac{2\hat{I}_{\rm NS}}{\alpha^3\beta^3\tilde{r}^3}\ .
\end{eqnarray}
Solve $\alpha^3$ from Equation (\ref{eq:beta}), then substitute it in Equation (\ref{eq:alpha}).
It reads
\begin{eqnarray}
    \frac{2M}{\beta}&=& \frac{(2\hat{I}_{\rm NS})^{1/3}}{(\beta^3-\beta^2)^{1/3}} - \frac{2\hat{I}_{\rm NS}}{\tilde{r}^2(\beta^3-\beta^2)}\ .
\end{eqnarray}
Reorganize the above equation as
\begin{eqnarray}
    \frac{2\hat{I}_{\rm NS}}{\tilde{r}^2}&=& (2\hat{I}_{\rm NS})^{1/3} (\beta^3-\beta^2)^{2/3} - 2M (\beta^2-\beta)\ ,
\end{eqnarray}
and differentiate both sides obtain coefficients of $d\tilde{r}$ and $d\beta$, respectively.
We can written $d\beta/d\tilde{r}$ in the following form
\begin{eqnarray}
    \frac{d\beta}{d\tilde{r}} = - \frac{4\hat{I}_{\rm NS}}{\tilde{r}^3}\ \left[ \frac{2}{3} \frac{(2\hat{I}_{\rm NS})^{1/3}}{(\beta^3-\beta^2)^{1/3}} (3\beta^2-2\beta) - 2M(2\beta-1) \right]^{-1},
\end{eqnarray}
The expression of $\beta_{\tilde{x}}$ is then expressed by
\begin{eqnarray}
    \partial_{\tilde{x}} \beta&=& \frac{d\beta}{d\tilde{r}}\ \frac{d\tilde{r}}{d\tilde{x}}\ ,
\end{eqnarray}
where
\begin{eqnarray}
    \frac{d\tilde{r}}{d\tilde{x}}&=& \frac{1}{\Omega} \frac{\tilde{x}}{\sqrt{\tilde{x}^2+\tilde{z}^2}}\ .
\end{eqnarray}

We then use $d\beta/d\tilde{r}$ to express $\partial_{\tilde{x}} \mathcal{F}$ and $\partial_{\tilde{x}} \mathcal{G}$,
which are coefficients in the L'H\^opital's rule to treat the LS.
Here we present the trivial derivation. Firstly, we need to calculate $d\alpha/d\tilde{r}$.
Multiply by $\alpha$, Equation (\ref{eq:alpha}) becomes
\begin{eqnarray}
    \alpha^3-\alpha&=& - \frac{2M}{\beta\tilde{r}}\ .
\end{eqnarray}
We obtain $d\alpha/d\tilde{r}$ by differentiating the both sides
\begin{eqnarray}
    \frac{d\alpha}{d\tilde{r}}&=& \frac{1}{3\alpha^2-1}\ \left[ \frac{2M}{\beta\tilde{r}^2} + \frac{2M}{\beta^2\tilde{r}} \frac{d\beta}{d\tilde{r}} \right]\ .
\end{eqnarray}
The differentials of coefficients $\mathcal{B}$ and $\mathcal{C}$ are written in
\begin{eqnarray}
    \partial_{\tilde{x}} \mathcal{B}&=& \left( \frac{\partial\mathcal{B}}{\partial\alpha} \frac{d\alpha}{d\tilde{r}} +  \frac{\partial\mathcal{B}}{\partial\beta} \frac{d\beta}{d\tilde{r}} \right)\ \frac{d\tilde{r}}{d\tilde{x}}\ ,  \nonumber\\
    \partial_{\tilde{x}} \mathcal{C}&=& \left( \frac{\partial\mathcal{C}}{\partial\alpha} \frac{d\alpha}{d\tilde{r}} +  \frac{\partial\mathcal{C}}{\partial\beta} \frac{d\beta}{d\tilde{r}} \right)\ \frac{d\tilde{r}}{d\tilde{x}}\ .
\end{eqnarray}
Re-write the coefficient $\mathcal{F}$ as
\begin{eqnarray}
    \mathcal{F}&=& \beta^{-2} (\mathcal{B} \tilde{x}^2 + \mathcal{C}) \left[ 1 - \frac{\tilde{z}^2}{\tilde{x}^2+\tilde{z}^2}  \right]\ .
\end{eqnarray}
The the differential $\mathcal{F}_{\tilde{x}}$ is calculated by
\begin{eqnarray}
    \partial_{\tilde{x}} \mathcal{F}&=& \beta^{-2} (\mathcal{B} \tilde{x}^2 + \mathcal{C}) \ \frac{2\tilde{x} \tilde{z}^2}{\left( \tilde{x}^2+\tilde{z}^2 \right)^2}  \nonumber\\
    &\ & +\ \beta^{-2} ( 2\tilde{x}\mathcal{B} + \partial_{\tilde{x}} \mathcal{B}\ \tilde{x}^2 + \partial_{\tilde{x}} \mathcal{C} )\ \frac{\tilde{x}^2}{\tilde{x}^2+\tilde{z}^2}  \nonumber\\
    &\ & -\ \frac{2\partial_{\tilde{x}} \beta}{\beta^3} (\mathcal{B} \tilde{x}^2 + \mathcal{C})\ \frac{\tilde{x}^2}{\tilde{x}^2+\tilde{z}^2}\ .
\end{eqnarray}
Similarly, the the differential $\mathcal{G}_{\tilde{x}}$ is calculated by
\begin{eqnarray}
    \partial_{\tilde{x}} \mathcal{G}&=& - \beta^{-2} (\mathcal{B} \tilde{x}^2 + \mathcal{C}) \ \frac{2\tilde{x} \tilde{z}^2}{\left( \tilde{x}^2+\tilde{z}^2 \right)^2}  \nonumber\\
    &\ & +\ \beta^{-2} ( 2\tilde{x}\mathcal{B} + \partial_{\tilde{x}} \mathcal{B}\ \tilde{x}^2 + \partial_{\tilde{x}} \mathcal{C} )\ \frac{\tilde{z}^2}{\tilde{x}^2+\tilde{z}^2}  \nonumber\\
    &\ & -\ \frac{2\partial_{\tilde{x}} \beta}{\beta^3} (\mathcal{B} \tilde{x}^2 + \mathcal{C})\ \frac{\tilde{z}^2}{\tilde{x}^2+\tilde{z}^2}\ .
\end{eqnarray}
Given $\partial_{\tilde{x}} \beta$, $\partial_{\tilde{x}} \mathcal{B}$, and $\partial_{\tilde{x}} \mathcal{C}$
derived above, the values of $\partial_{\tilde{x}} \mathcal{F}$ and $\partial_{\tilde{x}} \mathcal{G}$
on each grid at LS can be directly obtained.

% Don't change these lines
%\bsp	% typesetting comment
\label{lastpage}
\end{document}